\documentclass[article]{aa}  
\usepackage{epsfig}
\usepackage{natbib}
\bibpunct{(}{)}{;}{a}{}{,}
\usepackage{graphicx}
\usepackage{amsmath}
\usepackage{amssymb}
\usepackage{mathrsfs}
\usepackage{array}
\usepackage{listings}

\def\m2s2{\hbox{\,m$^{2}$\,s$^{-2}$}} 
\def\Msun{\hbox{$M_{\odot}$}}             

\def\P~{\hbox{$\tilde{P}$}}
\def\F-{\hbox{${\overrightarrow F}$}}

\def\exo1{\mbox{CoRoT-Exo-1}}

\def\om{\omega}

\begin{document}

\title{Structure, stability, and evolution of 3D Rossby vortices in protoplanetary disks }

\author{ S. Richard \inst{\ref{inst1}} \and P. Barge \inst{\ref{inst1}} \and S. Le Diz\`es \inst{\ref{inst2}} }
\author{ S. Richard \inst{1} \and P. Barge \inst{1} \and S. Le Diz\`es \inst{2} }

\institute{Aix Marseille Universit\'e, CNRS, LAM (Laboratoire d'Astrophysique de Marseille) UMR 7326, 38 rue F. Joliot-Curie, F-13388 Marseille, Cedex 13, France   \label{inst1} 
\and Aix Marseille Universit\'e, CNRS, Centrale Marseille, IRPHE (Institut de Recherche sur les Ph\'enom\`enes Hors Equilibre) UMR 7342, 49 rue F. Joliot Curie, F-13013 Marseille, France  
\\ \email{samuel.richard@oamp.fr, pierre.barge@oamp.fr, ledizes@irphe.univ-mrs.fr} \label{inst2}  }

\date{Received 28 june 2013 / Accepted 13 September 2013}

\abstract
{Large-scale persistent vortices could play a key role in the evolution of protoplanetary disks, particularly in the dead zone where no turbulence associated with a magnetic field is expected. These vortices are known to form easily in 2D disks via the Rossby wave or the baroclinic instability. In three dimensions, however, their formation and stability is a complex problem and still a matter of debate.}
{We study the formation of vortices by the Rossby wave instability in a stratified inviscid disk and describe their 3D structure, stability, and long-term evolution.}
{Numerical simulations were performed using a fully compressible hydrodynamical code based on a second-order finite volume method. We assumed a perfect-gas law and a non-homentropic adiabatic flow.} 
{The Rossby wave instability is found to proceed in 3D in a similar way as in 2D. Vortices produced by the instability look like columns of vorticity in the whole disk thickness; the weak vertical motions are related to the weak inclination of the vortex axis that appear during the development of the RWI. 
Vortices with aspect ratios higher than 6 are unaffected by the elliptical instability. They relax into a quasi-steady columnar structure that survives hundreds of rotations while slowly migrating inward toward the star at a rate that reduces with the vortex aspect ratio. Vortices with 
a lower aspect ratio are by contrast affected by the elliptic instability. Short aspect ratio vortices ($\chi <4$) are completely destroyed in a few orbital periods. Vortices  with an intermediate aspect ratio ($4<\chi < 6$) are partially destroyed by the elliptical instability in a region away from the midplane 
where the disk stratification is sufficiently strong. }  
{Elongated Rossby vortices can survive many orbital periods in protoplanetary disks in the form of vorticity columns. They could play a significant role in the evolution of the gas and the gathering of solid particles to form planetesimals or planetary cores, a possibility that receives a renewed interest with the recent discovery of a particle trap in the disk of Oph IRS 48. }


\titlerunning{Structure, stability, and evolution of 3D Rossby vortices in protoplanetary disks  }

\authorrunning{Richard et al. }
\maketitle

\section{Introduction}
\label{sec:intro}
The potential role of vortices in protoplanetary disks has been pointed out by \cite{Barge1995} pointed out the potential role of vortices in protoplanetary disks : they can capture large amounts of solid particles and can thus participate in the process of planet formation by speeding up the formation of planetesimals and the rapid growth of planetary cores. The formation, stability, and evolution of these vortices in protoplanetary disks have since raised a number of challenging questions.

Most numerical  studies of protoplanetary disks have been performed in 2D using quantities averaged over the disk thickness. 
 For  Keplerian disks, numerical simulations have shown that vortices can be created if 
the disk possesses some heterogeneities \citep{Bracco99}.
These are created at the border of the dead zone of protoplanetary disks \citep{Varniere2006}, for instance, or at the edge of a gap opened by a sufficiently massive planet 
\citep{devalborro2007}.
They can make the disk unstable with respect  to the  Rossby wave instability (thereafter RWI).  This instability  was first studied by  \cite{Lovelace1999} and \citet{Li2000, Li2001} in the context of the protoplanetary disks
and was proposed as a way to produce large-scale vortices. The RWI is a global, linear, and non-axisymmetric instability that occurs when an extremum of potential vorticity exists in the disk. 2D vortices can also be  produced by a non-linear baroclinic instability, as explored by \citet{Klahr2003, Johnson2005, Petersen2007a, Petersen2007b} and \citet{Lesur2010}. 
These vortices are known to migrate inward toward the star \citep{Li2000, Paardekooper2010, Surville2012, Surville2013}. 

The dynamics of the disk in three dimensions is a much more complex problem that has been addressed by a few authors only.
 The creation of vortices in 3D by itself is a problem.
In homentropic disks, \citet{Meheut2010,Meheut2012b} showed that the RWI could lead to the formation of complex 3D vortical structures with strong vertical motions.
The RWI was studied in more detail both theoretically and numerically for locally isothermal \citep{Lin2012} and non-barotropic disks \citep{Lin2013}.
Lin provided explanations for the different observations by \citet{Meheut2010,Meheut2012b}. He also showed that the most unstable mode of the RWI   
can remain quasi-2D in a non-homentropic disk.   Our simulations for a non-homentropic disk show that stable quasi-2D  vortices can 
indeed be created by the RWI. 

The vortices  may exhibit a complex dynamics.  \citet{Barranco2005} were the first to show that the dynamics of 2D vortices can be completely different in 3D, using anelastic numerical simulations in vertically stratified disks. 
Due to the strong shear they support, 2D vortices in a disk could a priori be unstable with respect to the 3D elliptical instability  \citep{Kerswell2002}.  This instability generically affects strained vortices \citep{Pierrehumbert86,  Bayly86} and is modified by the stratification and the background rotation of the disk \citep{Miyazaki93, Guimbard2010}.  
\citet[thereafter LP09]{Lesur2009} have considered the specific case of  the vortices in a Keplerian disk. Using simple vortex models in a shearing box, they  analyzed the stability of the vortices according to their aspect ratio in  a weak and moderately stratified case. Interestingly, 
they showed that  a moderate stratification tends to destabilize vortices that were stable without stratification.   

3D studies are also important to better understand the concentration of the solid particles in the vortices. Indeed, the preliminary two-phase simulations we performed showed that the dust-to-gas ratio depends strongly on height in the disk due to the vertical component of the Sun's gravity. A comprehensive study of the evolution of protoplanetary disks will be investigated with 3D two-phase simulations in a future work, in continuation of a preliminary 2D approach \citep{Inaba2006}.

In this paper, we are interested in the formation of 3D vortices by the RWI and in their long-term evolution. 
The basic equations, the disk model, and the numerical method are presented in Sect. \ref{equation}. Sect. \ref{sec:rossby} is devoted to the description of the linear growth 
 and non-linear evolution of the RWI. The characteristics of the final vortex, that are obtained after a succession of merging processes, are also provided
 in this section.
 The stability properties of the vortex with respect to the elliptical instability are examined 
 in Sect. 4, together with its property of migrating toward the star. 
 A brief conclusion is  provided in Sect. 5.

\section{Fluid equations and numerical method}
\label{equation}

\subsection{Standard disk assumptions}

The gas of the nebula was assumed to be a mixture of molecular hydrogen (75\%) and helium (25\%) with a mean molecular weight $\mu = 2.34 $g/mol. The low-pressure conditions inside the disk justify using perfect-gas law as the appropriate equation of state.  At steady-state the gas is stratified with an hydrostatic equilibrium in the vertical direction and a pressure-modified centrifugal equilibrium in the radial direction. 

As commonly done in the modeling of circumstellar disk observations, the surface density and temperature of the gas were assumed to decrease as simple power-laws of the distance to the star as $r^{-p}$ and $r^{-q}$.  
The disk self-gravity was neglected and we focused on optically thick regions of the nebula where the coupling of the gas with the magnetic field is negligible due to weak ionization. The problem was addressed by numerically solving the full set of the 3D compressible hydrodynamical equations in cylindrical coordinates.

\subsection{Governing equations}
\label{sec:equations}

The standard equations for an inviscid gas flowing around a central gravitational potential reads
\begin{equation}
{ {\partial \rho} \over {\partial t} } + {\vec\nabla} \cdot (\rho {\vec V}) =0
 \label{gov-eq1}
 \end{equation}
\begin{equation}
{ {\partial \rho {\vec V}} \over {\partial t} } +{\vec \nabla} \cdot (\rho {\vec V}{\vec V}) + {\vec \nabla} P =   {\rho {\vec \nabla} \phi} 
 \label{gov-eq2}
 \end{equation}
\begin{equation}
{ {\partial \rho e} \over {\partial t} } +{\vec\nabla} \cdot ({\vec V}(\rho e +P)) = \rho {\vec V} \cdot {\vec \nabla} \phi 
 \label{gov-eq3}
 \end{equation}
\begin{equation}
\rho e = {P\over{\gamma -1}} + {1\over 2}\rho {\vec V}^{2},
 \label{gov-eq4}
 \end{equation}
where  $\rho$ and $P$  are the density and pressure of the gas, and $u$, $v$ and $w$ are the radial, azimuthal, and vertical components of the gas velocity ; $\rho e$ denotes the total specific energy of the gas with adiabatic index $\gamma = 1.4$ and $\phi=GM_{\star}/(r^2+z^2)^{3/2}$ is the gravitational potential. Because no heat transfer was assumed in the disk, the energy equation is equivalent to the isentropic relation
\begin{equation}
{d\over {dt}} S = 0 ,
\label{entropie}
\end{equation}
where $S=\ln(P/\rho^\gamma)$ is the entropy and $ {d/dt}  = {\partial / \partial t} + {\vec V}\cdot {\vec \nabla}$ is the Lagrangian derivative.

The equations were used in non-dimensional form thanks to the following normalization:
\begin{subequations}
  \begin{align}
    r & =r_o \tilde{r} ~~~~~~~~  z  =r_o \tilde{z} ~~~~~~~~~~  \vec{V}=v_o \vec{\tilde{V}}  \\
   \rho & =\rho_o \tilde{\rho} ~~~~~~~  P  =\rho_o v_o^2 \tilde{P} ~~~~~~~  T= {v_o^2 \over R} \tilde{T}
  \end{align}
\end{subequations}
where the $_o$ index refers to their values at $r_o$ from the star in the midplane:
$v_o=\sqrt{GM_\star}/r_o^{1/2}$, $\rho_o=\rho (r_o,z=0)$.  In the following,  the tilde over the dimensionless variables is dropped. 
With these new variables, equations  \eqref{gov-eq1} - \eqref{gov-eq4}  are unchanged except for the gravitational potential, which reads $\phi=1/(r^2+z^2)^{3/2}$.  The important parameter is then  the Mach number $M_{A_o} = v_o/\sqrt{\gamma R T_o}$, where $T_o$ is the 
temperature computed at $r_o$ from the star.

\subsection{Stable equilibrium of the disk}

In the absence of perturbation, the gas is in a steady-state equilibrium, flowing at nearly the Keplerian velocity around the star.
The vertical and centrifugal equilibrium equations for the axisymmetric steady-state are
\begin{subequations}
  \begin{align}
{{\partial P}\over{\partial z}}=-{{\rho z} \over (r^2+z^2)^{3/2}} ~~,
\label{equ:Pz} \\
{{\partial P}\over{\partial r}}=-{{\rho r} \over (r^2+z^2)^{3/2}}  + \frac{\rho v^2}{r}~~.
\label{equ:Pr}
 \end{align}
\end{subequations}

If we assume that the temperature does not depend on the height $z$, we can write $T_D$ as 
\begin{equation}
T_D(r)=  \frac{r^{-q}}{\gamma M_{A_o}^2}.
\end{equation}
From Eq. (\ref{equ:Pz}) and 
\begin{equation}
P_D(r,z)  = \rho_D(r,z)~T_D(r),
\end{equation}
we then easily find the density 
\begin{equation}
\rho_D(r,z)  = r^{-\alpha} \exp{\left( {1 \over T_D(r)}\left({1 \over \sqrt{r^2+z^2}}-{1 \over r}\right)\right) }   \label{rhoD} .
\end{equation}
 Integrated over the disk thickness, \eqref{rhoD} gives the surface density
 \begin{equation}
\Sigma_D(r)  \ \approx \sqrt{2\pi \over \gamma}  {r^{-p} \over M_{A_o}},
\end{equation}
provided that $\alpha=p+(3-q)/2$. 
The pressure height is given by 
\begin{equation}
H_D(r) =\frac{ r^{(3-q)/2}}{M_{A_o}}.
\label{exp:HD}
\end{equation}

The azimuthal velocity deduced from the radial centrifugal equilibrium is
   \begin{equation}
  v_D(r,z)  = \sqrt{{1\over r}-\frac{3+2p+q}{2}T_D(r)+ q\left({1\over \sqrt{r^2+z^2}}-{1 \over r} \right) } ~ .
\end{equation}

This equilibrium solution only depends on three parameters: the indices $p$ and $q$, and the Mach number $M_{A_o}$. 
Note in particular that the disk structure does not depend on its mass. 
Except for the comparison with Lin's results, we took $q=0.5$, $p=1.5$, and $M_{A_o}=25$ throughout. 
This value of the Mach number is obtained for $r_o=1AU$, $T_o=280K$, and $M_\star = 1 \Msun$.

\subsection{Unstable equilibrium state}

\label{sec:initial} 

 The above solution is linearly stable.  To render the solution unstable with respect to the Rossby wave instability, 
we chose to add an annular bump in the density and pressure fields while keeping  the temperature profile unchanged,
\begin{subequations}
  \begin{align}
    \rho(r,z) & =\rho_D(r,z)(1+f(r)) \\
    P(r,z) & =P_D(r,z)(1+f(r)), 
        \end{align}
\end{subequations}
with 
 \begin{equation}
f(r)=A\exp\left(-\frac{(r-r_b)^2}{ \sigma^2}\right). 
\label{exp:f}
\end{equation}
The velocity field is then obtained from Eq. (\ref{equ:Pr}).
In a  protoplanetary disk, this density bump may be caused, for example, by an axisymmetrical accumulation 
of gas at the edge of the dead zone. 
The Rossby wave instability is characterized in 2D 
by the function ${\cal L}$   defined by 
\begin{equation}
{\cal L} = \frac{\Pi ^{2/\gamma_2} }{2\Sigma \zeta},
\end{equation}
where $\gamma_2 $ is the adiabatic index, and $\Pi$ and $\zeta$  are the vertically integrated pressure and {total} axial vorticity.
Following \citet{Lin2013}, we assumed the relation $\gamma_2 =(3\gamma -1)/(\gamma +1) \approx  1.33$ obtained
by \cite{Goldreich1986}.

\begin{figure}
\begin{center}
\includegraphics[width=9cm]{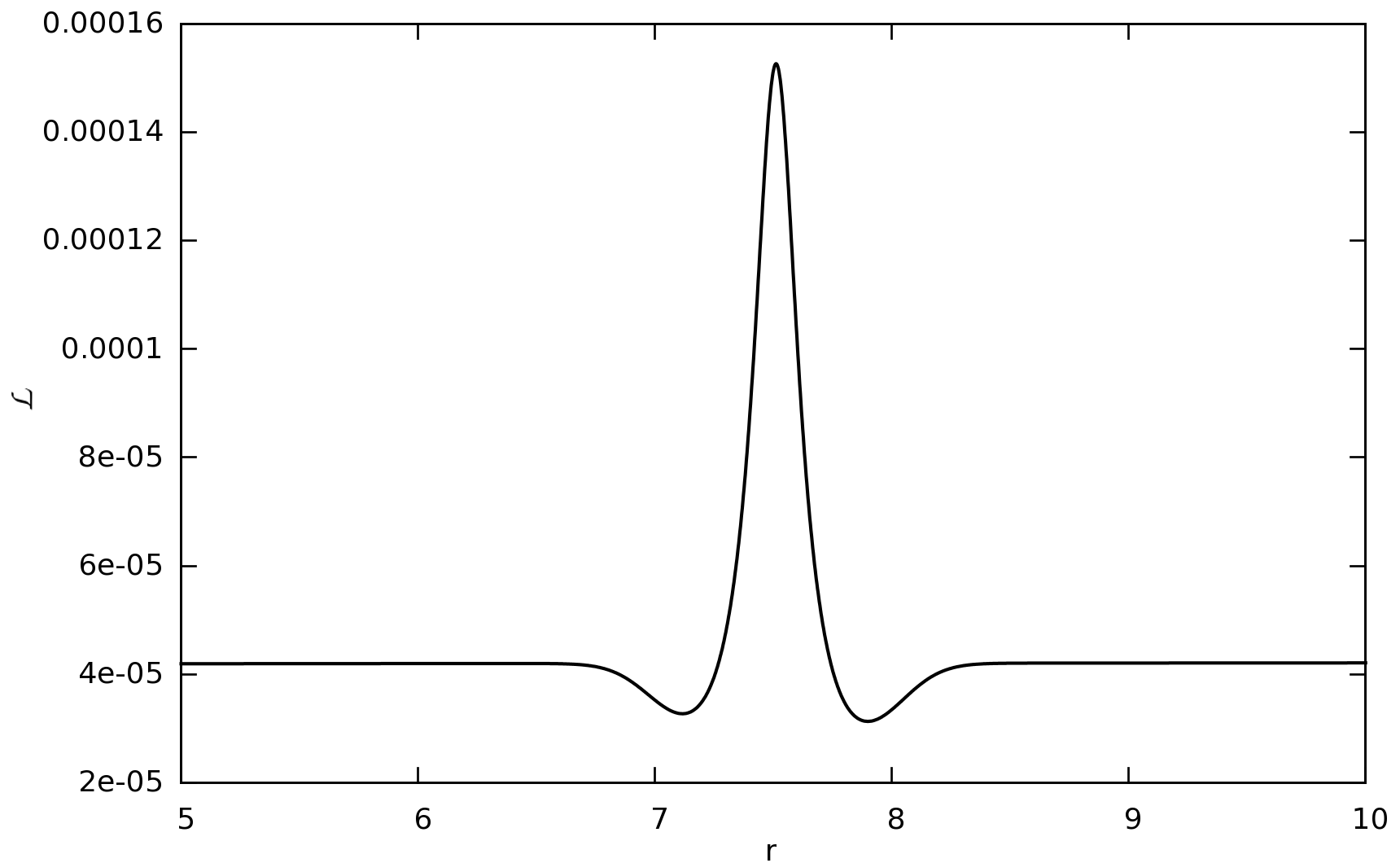}
\includegraphics[width=9cm]{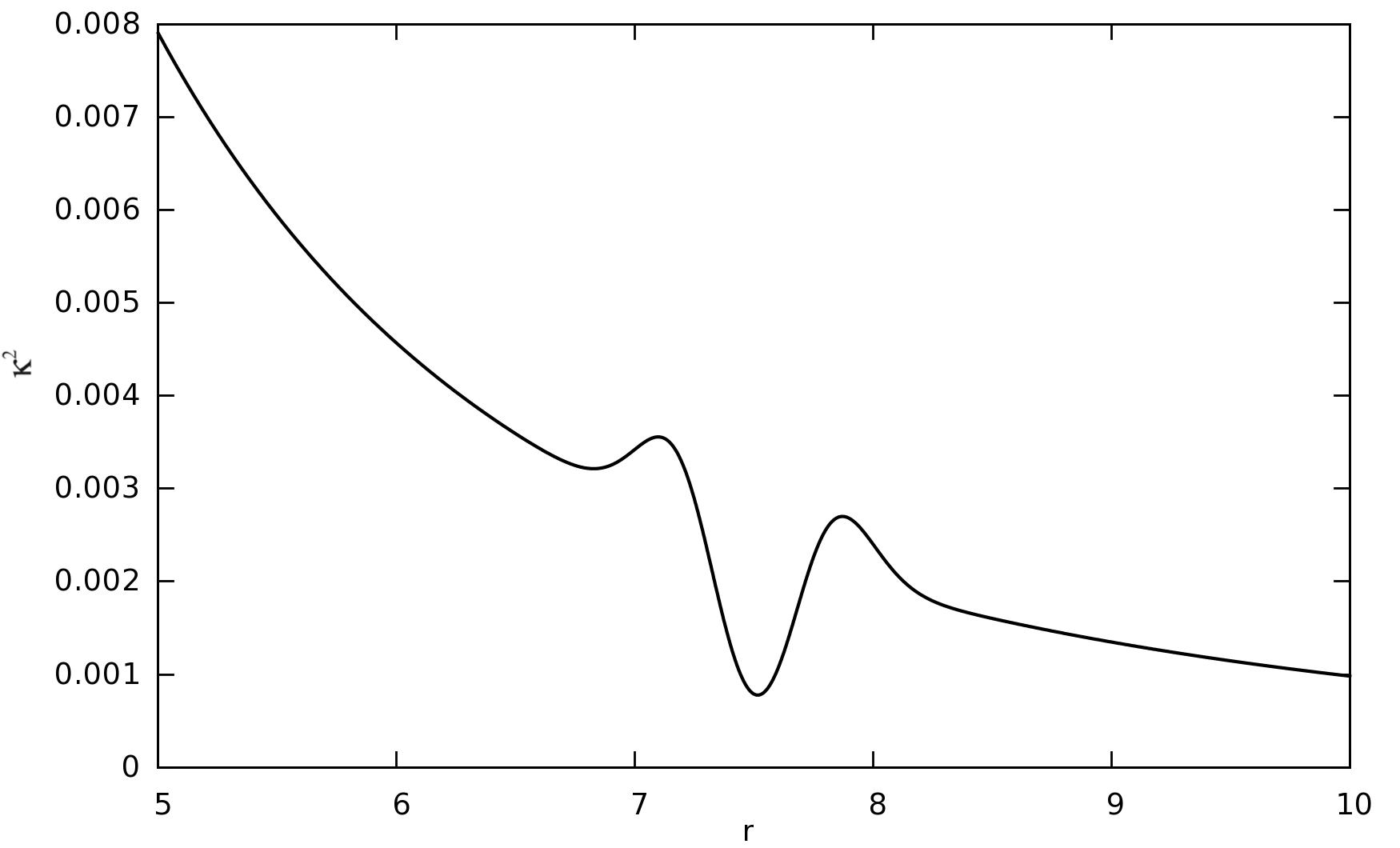}
\end{center}
 \caption{Radial dependance of the function ${\cal L}$ (top) and the function $\kappa^2_0$ (bottom) that characterize the Rossby wave instability and
 the centrifugal instability. Disk parameters are $q=0.5$, $p=1.5$, $M_{A_o}=25$, $\sigma = 0.3$, $r_b=7.5$, and $A=0.2$.}
\label{fig:LKappa}       
\end{figure}

As shown by Lovelace et al (1999), the  function ${\cal L}$ must exhibit a local maximum for instability.   
  {The centrifugal instability is characterized by the function  $\kappa^2 ={ 2 \Omega \over r}{d \over dr}  r^2 \Omega $, 
which becomes negative  in the presence of centrifugal instability.  
The function $\kappa^2$ weakly depends on $z$. In the following, we consider the value in the midplane ($z=0$), where the absolute value of $\kappa^2$ is maximum, and denote this value by $\kappa_0^2=\kappa^2(z=0)$.  
For $A=0.2$, $r_b=7.5$ and $\sigma=0.3$, we obtain the typical plots shown in Fig.~\ref{fig:LKappa}
for the functions ${\cal L}$ and $\kappa_0 ^2$. 
The function ${\cal L}$ does exhibit a local maximum, while $\kappa_0^2$ remains positive.
For higher values of $A$, $\kappa_0^2$ can become negative, which implies centrifugal instability. For $\sigma=0.3$, $r_b=7.5$, the threshold for centrifugal instability is $A_c=0.334$.
Except in Sect. \ref{sec:stability}, $A$ was chosen to be sufficiently small to remain in the centrifugally stable regime.}

The disk is stably stratified. The stratification is
characterized by the Brunt-V\"ais\"al\"a frequency,
\begin{equation}     
N^2={{z\over (r^2+z^2)^{3/2}} \left({1\over \gamma P_D} {\partial P_D \over \partial z}- {1\over \rho_D} {\partial \rho_D \over \partial z} \right )}  ~~~.
 \label{N2-1}
 \end{equation}
 In our case, both the stable and unstable equilibrium states have a 
 Brunt-V\"ais\"al\"a frequency given by
 \begin{equation}
 N^2={z^2\over (r^2+z^2)^{3}}{\gamma-1 \over \gamma T_D}   ~~~.
 \label{N2-2}
 \end{equation}
  This function vanishes in the equatorial plane, and increases with height until it reaches a maximum at $z=\sqrt{2}r$.

\subsection{Numerical method and validation}
The system of non-linear equations was solved using the finite-volume method. We used a second-order finite-volume scheme, the MUSCL Hancock scheme, and an exact Riemann solver. 
The present 3D version of the code was issued from a 2D version described in a previous paper \citep{Inaba2005}, which was recently improved with a well-balanced numerical scheme  that enables one to preserve the stationary solution over very many rotation periods. This scheme was developed in 2D by C. Surville (unpublished report) and is implemented in the 3D version of the code used here.

The computational domain for our simulations was defined as 
$5 < r < 10 $ , $0<\theta<2\pi$ and $0<z<1$ (equivalent to 2.4 scale heights for a ring centered at $r=7.5$). 
The mesh was a regular one with 200 cells in the radial direction, 300 in the azimuthal direction and 100 in the vertical direction 
{which, per scale height, corresponds to 20 cells in the radial direction, 3 cells in the azimuthal direction and 50 in the vertical direction (priority was given to the $z$ direction for studying the vertical effects).} 
The boundary conditions were imposed using two ghost cells at each boundary. 
In the azimuthal direction periodic conditions were used to be consistent with a global disk geometry. 
In the radial direction steady-state values were imposed on the ghost cells to connect the computational domain with the rest of the disk; 
these conditions are relatively permeable to perturbations of the steady-state and avoid spurious wave reflexions. 
In the vertical direction, boundary conditions were symmetric at level $z=0$ and extrapolated at $z= 1$, 
{ imposing zero vertical derivatives for the relative velocity and for the density and pressure ratios:}
\begin{equation}
    {\partial \over \partial z} (\vec {V} - \vec{V_D})=0 ~~~~~~~~~~ {\partial \over \partial z} {\rho \over \rho_D}=0 ~~~~~~~~  {\partial \over \partial z} {P \over P_D}=0,   
\end{equation}

{Other boundary condition were also tested, but turned out to be less efficient because of stronger residual perturbations.}

To test the stability of the code we first performed a number of long-term simulations starting from the stable stationary solution presented above. We found that after a hundred rotations, deviations from the initial solution are very weak and have no systematic trends: the radial and vertical velocities are lower than $10^{-12}$, the relative error on the density is smaller than $10^{-9}$. Therefore we are confident that our code can conserve the unperturbed steady-state solution even on the long runs made here. 

We also tried to reproduce one of the linear stability results obtained by \cite{Lin2013}.
\begin{figure}
\begin{center}
\includegraphics[width=9cm]{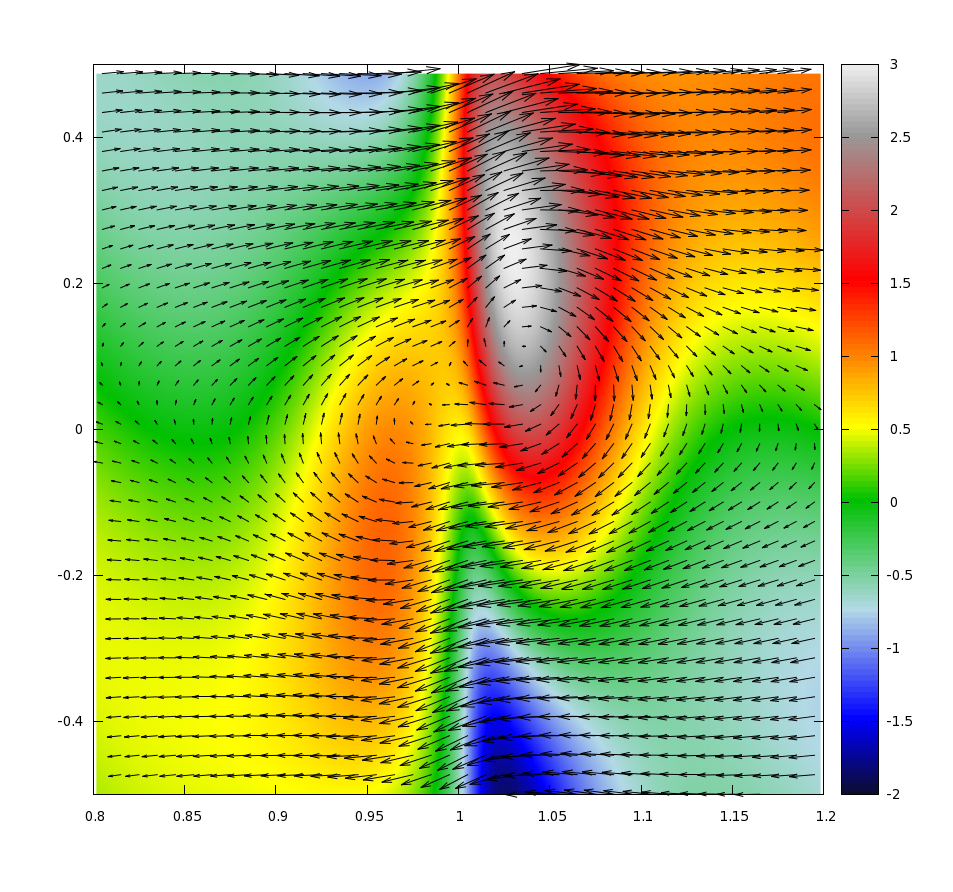}
 \caption{Density contour and velocity  field in the midplane ($r$, $\theta$) for the most unstable mode $m=4$ and for the parameters of \cite{Lin2013} $q=0$, $p=1.5$, $M_{A_o}=23.15$, $A=0.25$, and $\sigma=0.53$.}
\label{fig:densityLin}
\end{center}
\end{figure}
The parameters used by \cite{Lin2013}, \S 6.3 give in our notation $q=0$, $p=1.5$,  $\alpha=3$, $M_{A_o}=23.15$, and a Gaussian
bump with $A=0.25$ and $\sigma= 0.53 $.
Analyzing the time evolution of very weak perturbations, we were able to compute the linear growth rate of each azimuthal mode.  We found that the most amplified mode is for $m=4$ with a growth rate $\omega_i \approx 0.190$, which is close to the value $\om_i\approx 0.1937$ obtained by Lin.  In Fig.~\ref{fig:densityLin}, we displaye  the density field (colors) and the velocity field in the midplane (arrows) of the $m=4$ mode during the linear phase. When we compare this figure with the Fig. 18 (top) of \cite{Lin2013}, we see that the perturbation fields are very similar.
This confirms the good agreement between the two results and also validates our numerical code. 

\section{Formation and structure of Rossby vortices}
\label{sec:rossby}

In this section, we compute the non-linear evolution of the disk that results from 
introducing a small-amplitude white noise on the first five modes for the parameters given in \S  \ref{sec:initial}.   
{We monitored density, pressure, and velocity fields of the disk by considering relative quantities with respect to the stable equilibrium state.}
\begin{figure*}[t]
\begin{center}
\includegraphics[width=4cm]{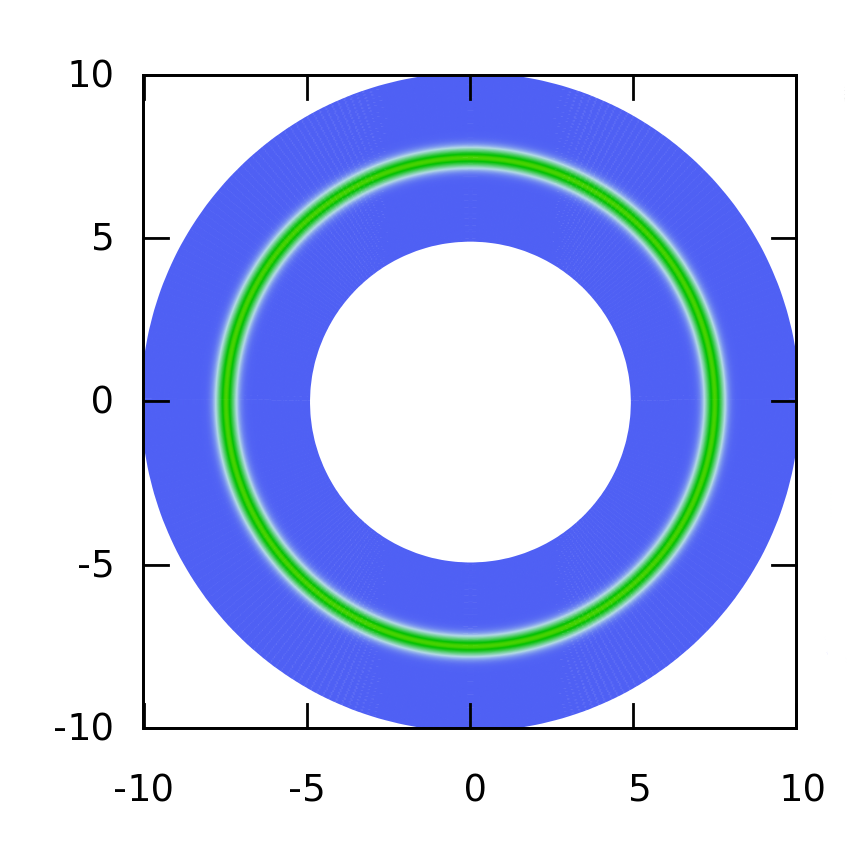}
\includegraphics[width=4cm]{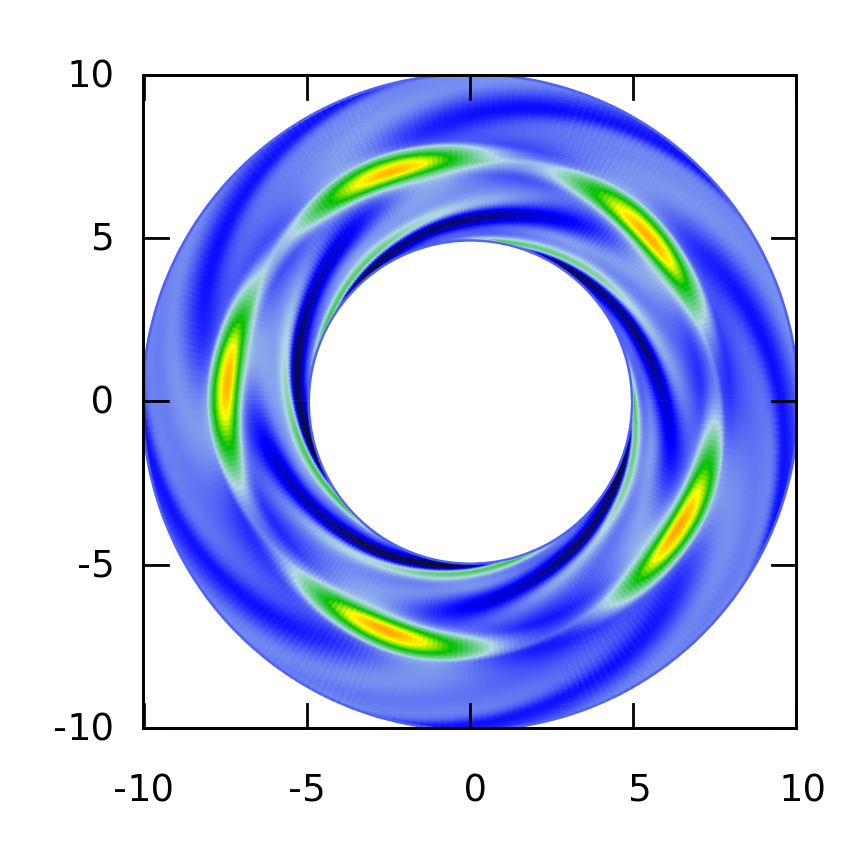}
\includegraphics[width=4cm]{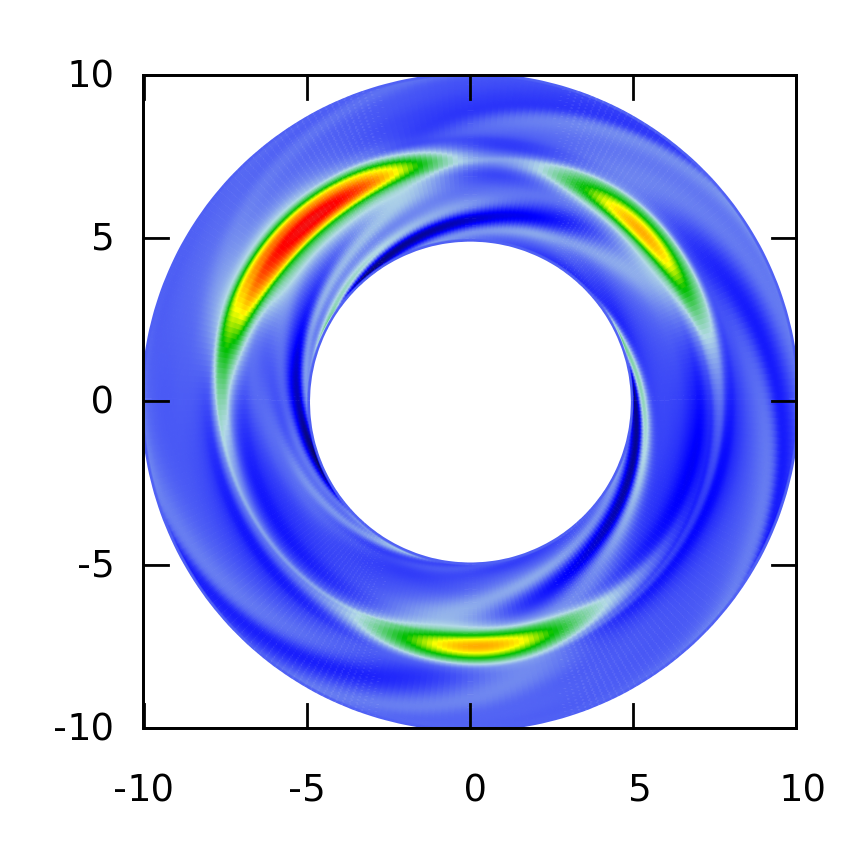}
\includegraphics[width=4cm]{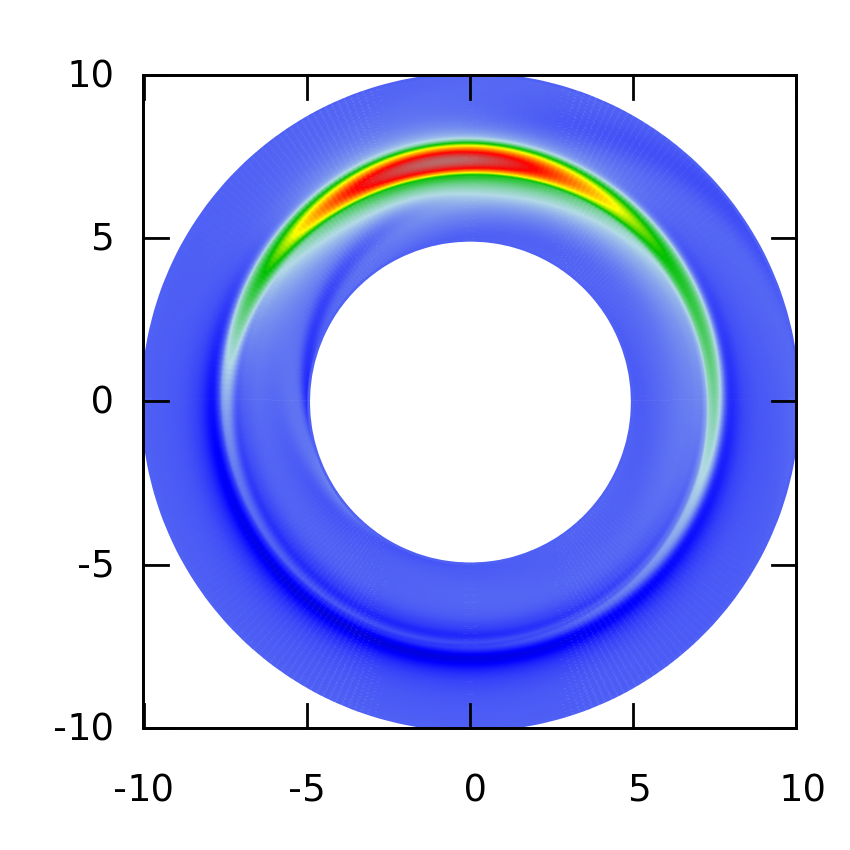}
\includegraphics[height=4cm]{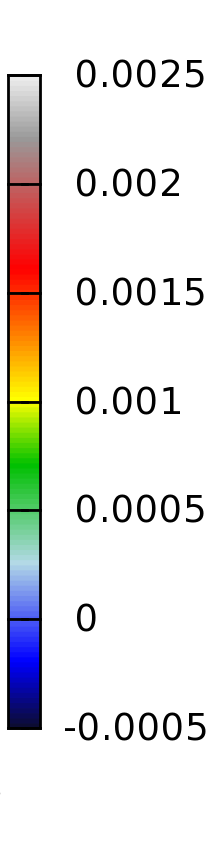}
\includegraphics[width=4cm]{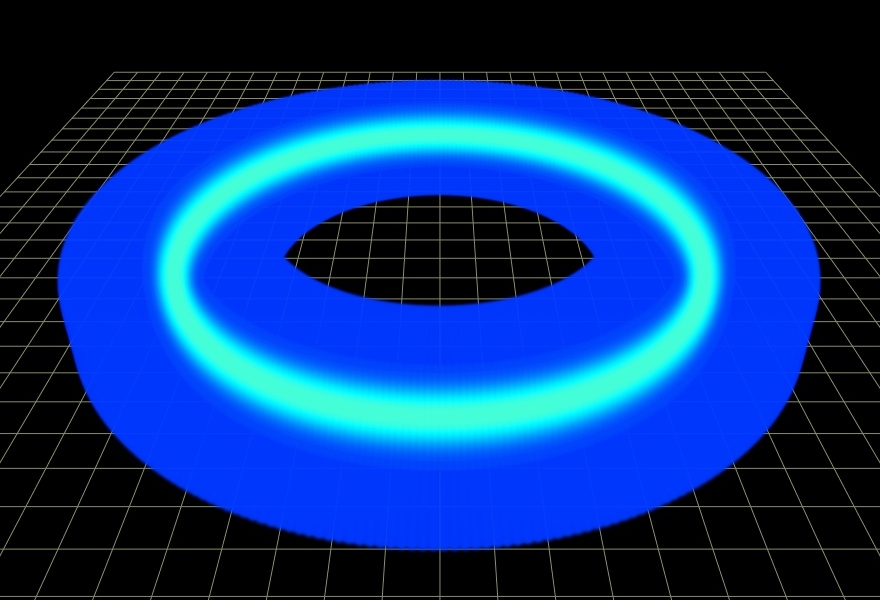}
\includegraphics[width=4cm]{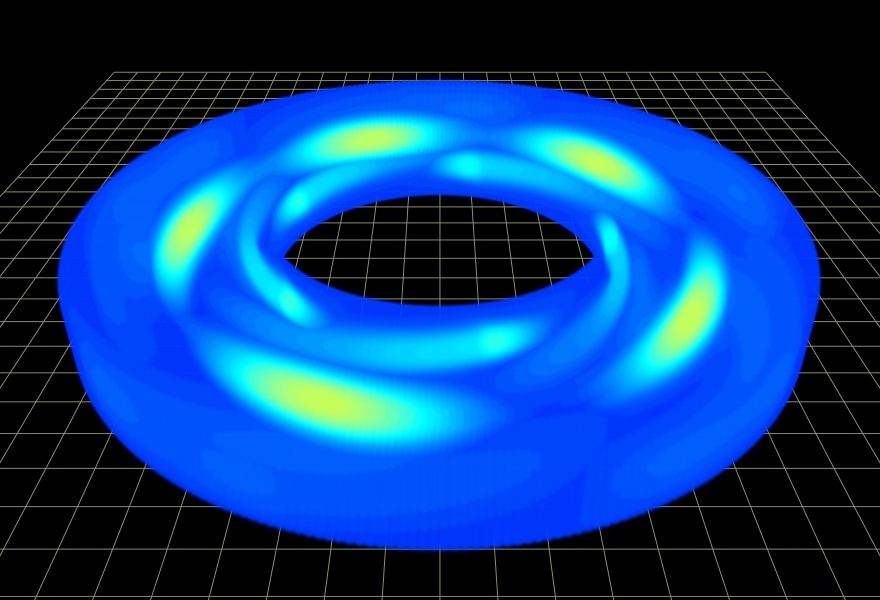}
\includegraphics[width=4cm]{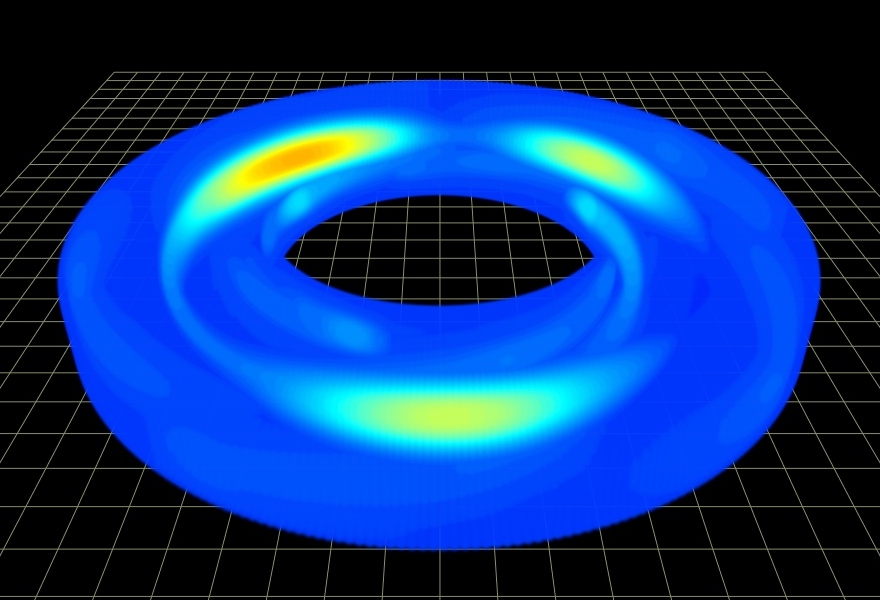}
\includegraphics[width=4cm]{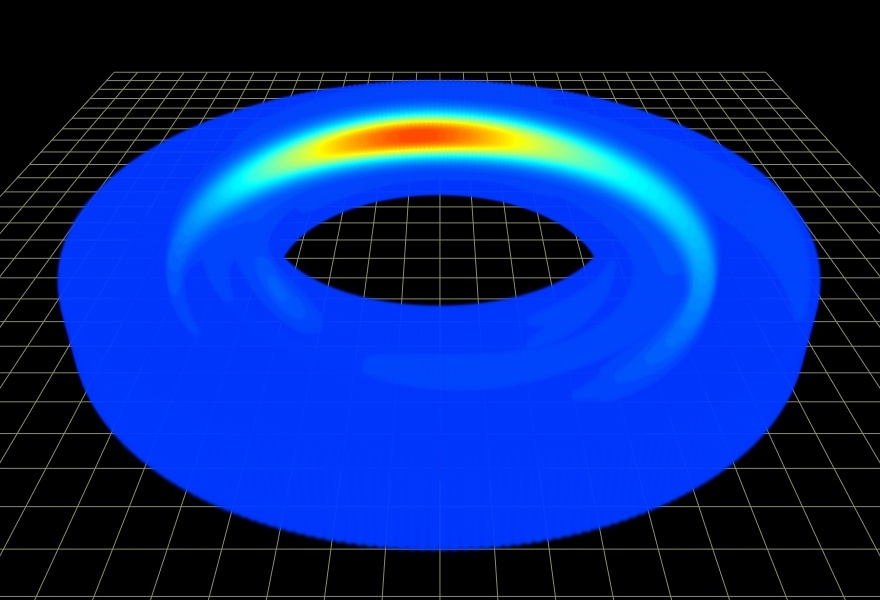}
\end{center}
 \caption{Evolution of the density ($\rho-\rho_D$) in a Rossby-unstable inviscid disk. Top: evolution in the $r - \theta$ plane ($z = 0$); bottom: evolution in a 3D perspective (visualisations performed with \textit{glnemo2}). From left to right, density values are plotted at initial time and after 20, 40, and 530 rotations.}
\label{Evolution}       
\end{figure*}
The 3D evolution of the density ($\rho-\rho_D$) is presented in Fig.~\ref{Evolution}. First, the instability grows linearly way and the annular overdensity fragments in a chain of vortices coupled to spiral waves that propagate on each side of the initial ring.  After the instability saturates, vortices tend to catch one another and successively merge into increasingly larger structures. After a few hundred rotations, a single vortex remains in the computational domain. This vortex persists for a long time (more than 600 rotations), slowly drifting toward the star. A slight decay is observed that corresponds to the unavoidable numerical diffusion. 
We now detail the different steps of this evolution.

\subsection{Growth and merging}

\begin{figure}
\begin{center}
\includegraphics[width=9cm]{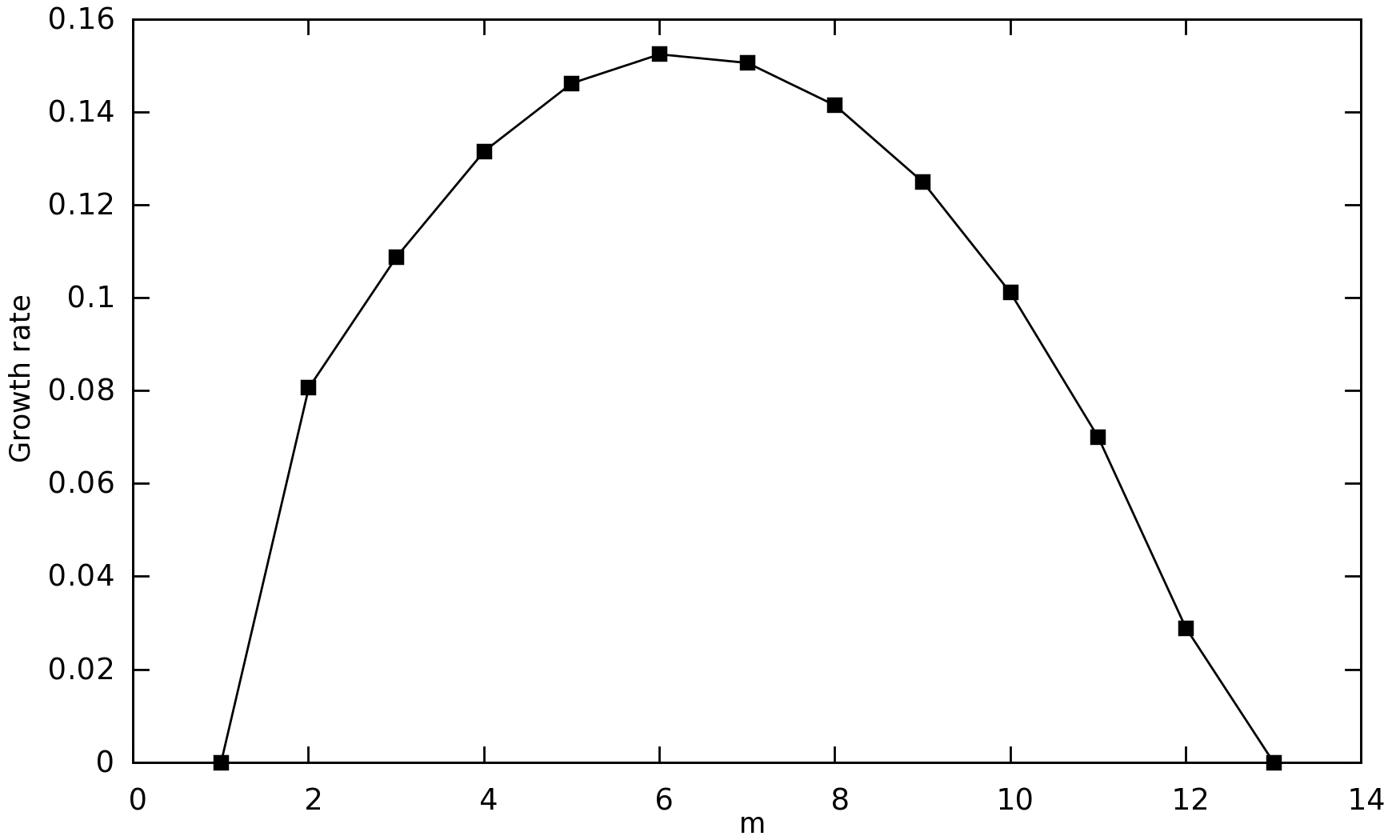}
 \caption{Growth rate of the most unstable mode versus the azimuthal wavenumber $m$.}
\label{fig:growthrate}
\end{center}
\end{figure}

\begin{figure}
\begin{center}
\includegraphics[width=9cm]{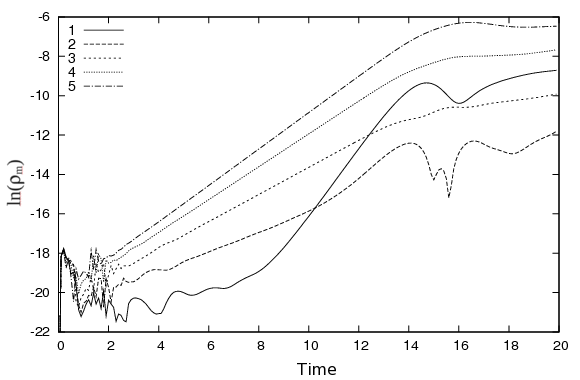}
 \caption{Time evolution of the amplitude of modes 1 to 5 during the first 20 rotations.}
\label{fig:growth}
\end{center}
\end{figure}

\begin{figure}
\begin{center}
\includegraphics[width=8cm]{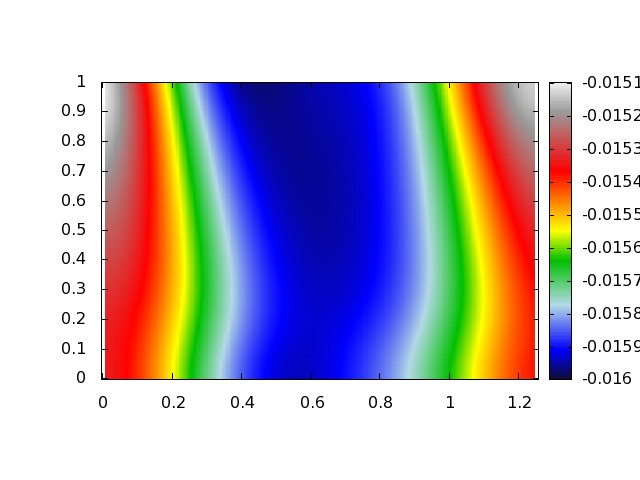} \\[-0.8cm]
\includegraphics[width=8cm]{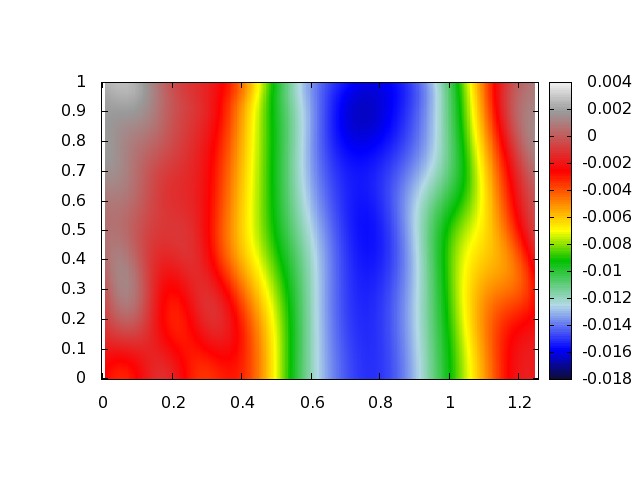} \\[-0.8cm]
 \caption{Vertical component of the vorticity in the $(\theta,z)$ plane after 11 rotations (top) and 22  rotations (bottom).}
\label{tilt}
\end{center}
\end{figure}

The first step in the development of the Rossby wave instability is a linear growth phase. The growth rate of the different instability modes is plotted in 
Fig.~\ref{fig:growthrate}. From this figure, we expect all the modes with $2\leq m \leq 12$ to possibly grow  during the linear phase. 
 Mode $m=6$ is expected to be the most unstable mode.  
In the simulation, where white noise was added to the first five modes, the growth of the  modes $m=2,3,4,$ and $5$ follows the linear prediction
(see Fig.~ \ref{fig:growth}). Mode $m=1$,  which is linearly stable,  starts to grow when the amplitude of the other modes has reached a 
sufficiently large amplitude. 
During the linear growth phase, the axial vorticity exhibits the form of  a tilted structure, as observed in Fig.~\ref{tilt} (top). This
agrees with the structure of the most unstable linear mode  
described by \cite{Lin2013}. 

{We point out that the vertical shear of the background flow cannot be the cause of the observed tilt. Indeed, tilted vorticity columns are also found in simulations with $q=0$, that is, without any vertical shear, and in the linear calculations performed by \citet{Lin2013}. 
No significant differences have been observed between simulations with vertical shear ($q\ne 0$) and simulations without vertical shear ($q=0$).}

Finally, after 22 rotations when the instability has saturated,  the 
tilt has almost completely disappeared (Fig.~\ref{tilt} bottom). This non-linear phenomenon is not observed in the linear framework of Lin (2013). Then the vortices evolve in the form of straight columns. 

The vortices are strongly deformed by the background shear. In the $(\theta,r) $ plane, the axial vorticity contours take the form of elongated ellipses
with an aspect of order 7. 
As soon as the vortices are formed, they start to interact with each other.  A first merging of two pairs of vortices is observed at 45 rotations, then a second
  merging process occurs   at  65 rotations,  before the last merging of the two remaining vortices after 500 rotations.  

\begin{figure}
\begin{center}
\includegraphics[width=9cm]{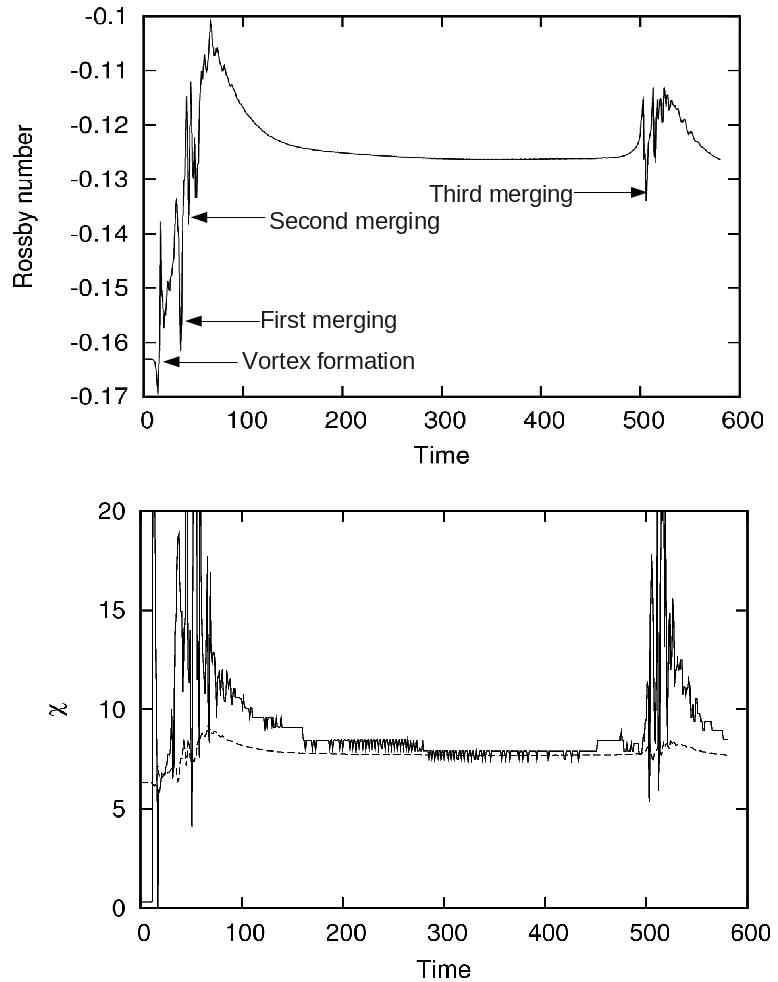}
 \caption{Time evolution of the main characteristics of the largest observed vortex. Top: Rossby number of the vortex as defined in (\ref{eq:Ro}); bottom: aspect ratio measured from the
 axial vorticity contours (solid line) and obtained from formula (\ref{exp:Kida}) (dashed line).}
\label{fig:aspectratio}
\end{center}
\end{figure}

{The Rossby number and the aspect ratio of the largest observed vortex were monitored during this evolution and are reported in Fig.~\ref{fig:aspectratio}. 
The Rossby number is defined by
\begin{equation}
Ro={\omega_{z_0} \over 2 \Omega_{D_0}},
\label{eq:Ro}
\end{equation}
where $\omega_{z_0}=\omega_z(0)$ is the relative axial vorticity of the vortex in the midplane  (i.e., the total vorticity in the midplane after removing of the background vorticity), and $\Omega_{D_0}$ the background angular velocity in the midplane at the vortex center.}

Despite the difficulty fo defining an aspect ratio during the merging processes, we observe that
the aspect ratio tends to increase. After the last merging process, the final vortex has an aspect ratio of close to 10.
This evolution of the aspect ratio agrees with the evolution of the norm of the axial vorticity maximum, which 
 tends to decrease. 
 For a vortex with uniform vorticity \citep{Kida81,CHavanis2000},  aspect ratio and vorticity are linked by the relation
 \begin{equation}
{\omega_{z_0} \over s} = -{1\over \chi}{\chi+1 \over \chi -1} ~,
\label{exp:Kida}
\end{equation}
where $s= -r_b \partial_r\Omega (r_b)$ is the local shear rate in the midplane at the position of the vortex.
We observe that this formula provides a good estimate of the aspect ratio of our non-uniform vortices 
if we take the maximum vorticity for $\omega _{z_0}$ in Eq. (\ref{exp:Kida}).   
 Note that the resulting vortices always have an aspect ratio higher than or equal to 7. This property can explain their 3D stability, as shown below. 
 
{\citet{Barranco2005} found that synthetic vortices are unstable under anti-symmetric perturbations. To examine this point a global simulation was performed without imposing the symmetry in the midplane. We obtained similar results. In particular, we did not observe the growth of any anti-symmetric perturbation.}
  
\subsection{Structure of the vortices}

\begin{figure*}
\centering
\includegraphics[width=7.cm, height=3.8cm]{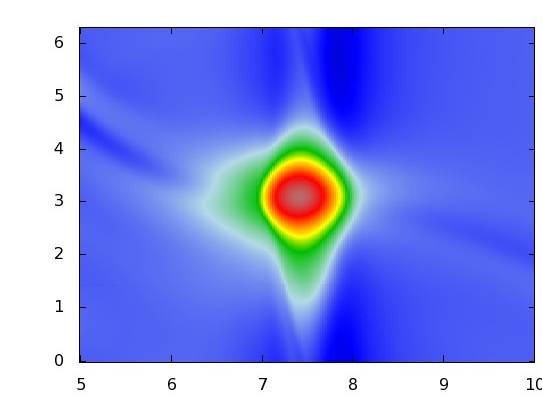}
\includegraphics[width=8.cm, height=3.8cm]{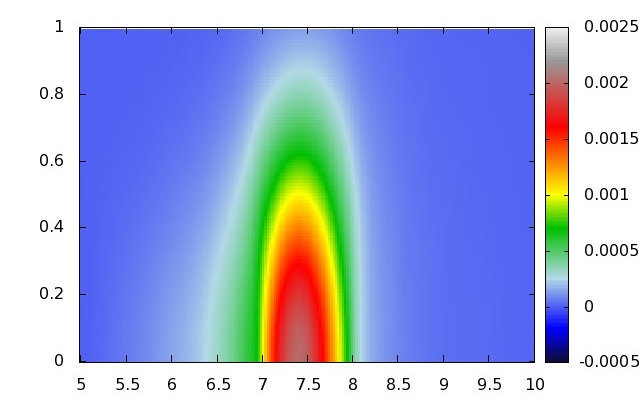}
\includegraphics[width=7.cm, height=3.8cm]{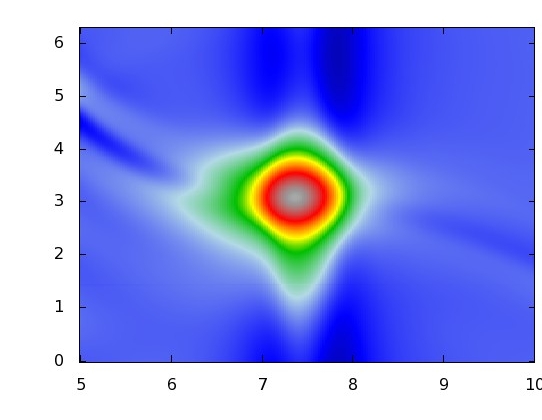}
\includegraphics[width=8.cm, height=3.8cm]{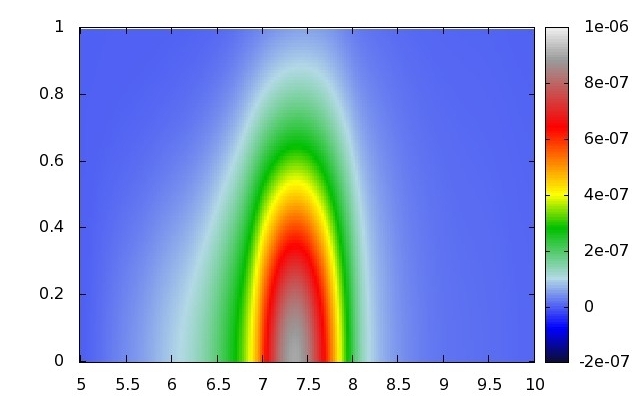}
\includegraphics[width=7.cm, height=3.8cm]{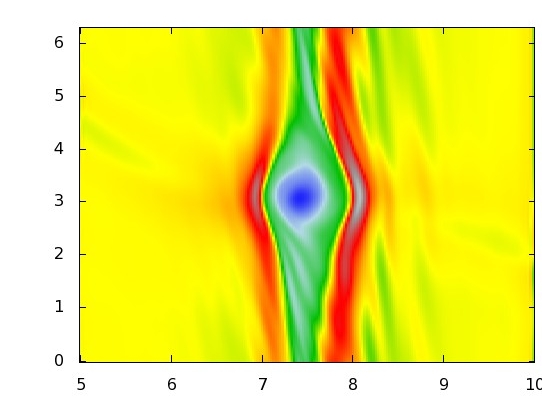}
\includegraphics[width=8.cm, height=3.8cm]{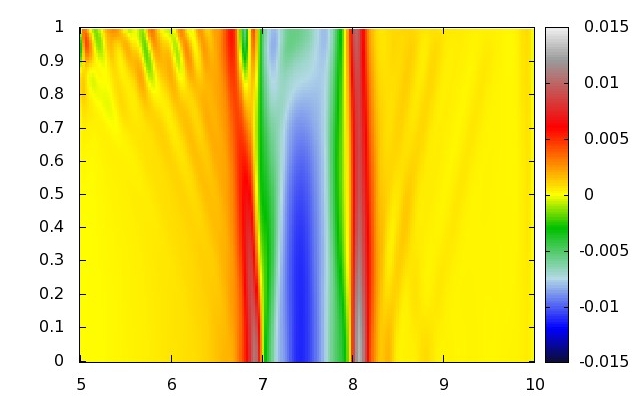}
\includegraphics[width=7.cm, height=3.8cm]{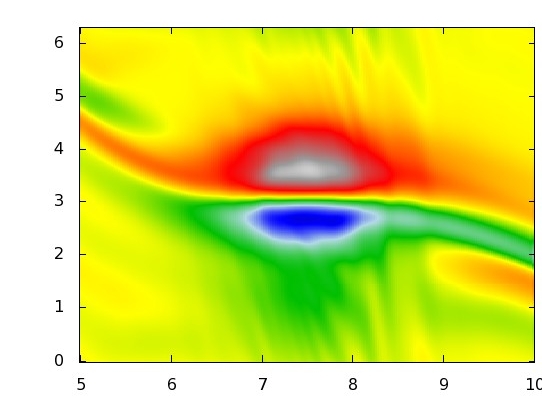}
\includegraphics[width=8.cm, height=3.8cm]{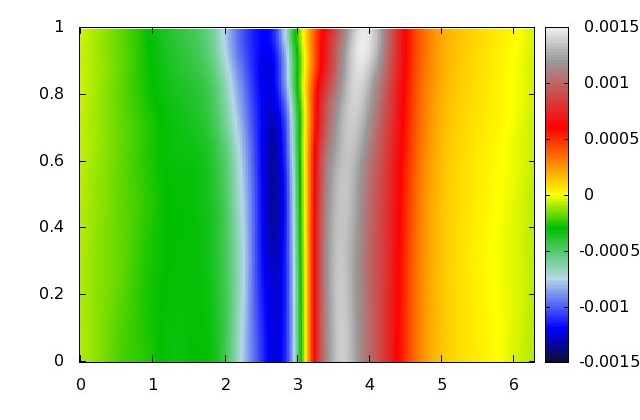}
\includegraphics[width=7.cm, height=3.8cm]{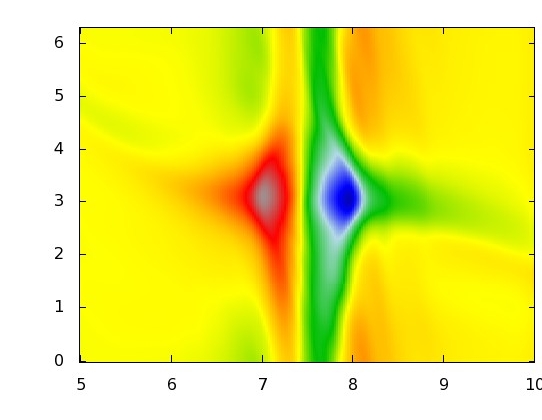}
\includegraphics[width=8.cm, height=3.8cm]{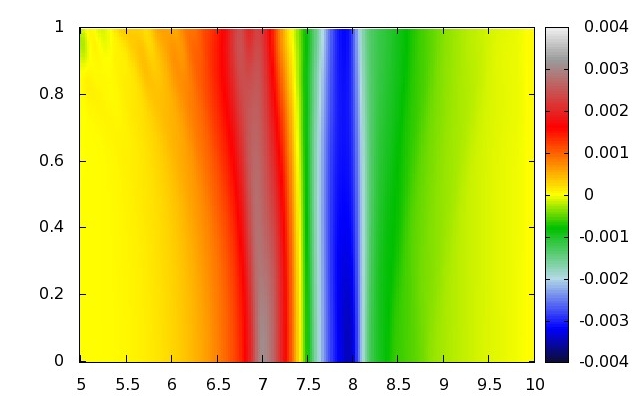}
\includegraphics[width=7.cm, height=3.8cm]{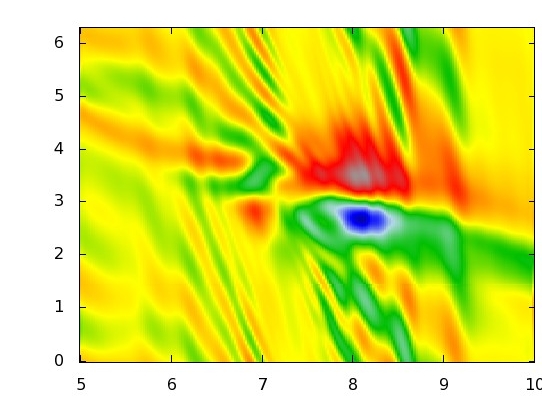}
\includegraphics[width=8.cm, height=3.8cm]{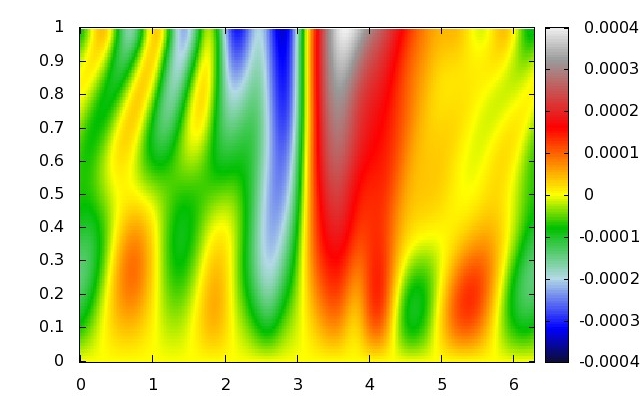}
\caption{Structure of a 3D Rossby vortex. From top to bottom the various rows show {  the fields (where the stable equilibrium state has been removed) of} density, pressure, axial vorticity, radial velocity, azimuthal velocity, and vertical velocity. The left column shows these values in the $r-\theta$ plane at z=0 except for the vertical velocity (last row), which is at z=0.5. The right column show the vertical profiles of the same values in the $r-z$ plane at $\theta=2.67$; radial velocity (fourth row) and vertical velocity (last row) are plotted in the $\theta-z$ plane at $r=8$.}
\label{structure}       
\end{figure*}

After the last merging process a single vortex remains in the simulated region of the disk. {This vortex is isolated, with no observed residual of the initial density bump, and can be characterized by a Rossby number  $Ro\simeq-0.12$ .} 
It is advected by the background flow and is found to form a quasi-steady solution of equations (1)-(4) in the rotating frame. 
The characteristics of this final vortex are given in 
Fig.~\ref{structure} after 530 rotations. 
The first five plots in the left column provide {the relative fields (where the stable equilibrium state has been removed) of} density, pressure, axial vorticity, and the radial and azimuthal velocity of the vortex
in the equatorial plane. They are typical  of an anti-cyclonic motion around a pressure and density  bump. 
These plots are very similar to that obtained with 2D simulations starting from identical initial conditions.
The axial dependence is shown in the right column of Fig.~\ref{structure}.
We observe that axial vorticity, and the radial and azimuthal velocity have a columnar structure in the whole thickness of the disk. 
The vorticity defects observed at the disk edge ($z=1$)  are weak imperfections due to the use of  not purely transparent boundary conditions.
Inside the vortex, the pressure and density fields are found to have a Gaussian profile like the background-disk state. 

In the equatorial plane, the axial velocity is null by symmetry.  
It is non-zero outside this plane but, takes values  one order of magnitude lower than the other velocity components. {In particular,  the highest Mach number of the vertical motions is lower than $0.02$}.  
This field is characterized in two perpendicular planes in the last row of Fig.~\ref{structure}. 
We observe higher values close to the vortex center, but the field is not as well localized. 
The highest axial velocity is so weak that 
we can even observe the trace of  gravity waves in the axial flow contours. These waves result from the stratification of the disk. 
Note, however, that  the axial velocity field has a structure taht resembles the radial velocity field.  
Therefore, the axial flow structure is most likely caused by a slight tilt of the vortex axis.

In conclusion, this 3D evolution of the Rossby wave instability is very similar to the 2D evolution. 
The 3D effects are weak perturbations that mostly result from the initial 3D structure of the background disk.   
A slight vortex tilting, which is characteristic of the linear RWI  in a non-homentropic disk \citep{Lin2013}, seems to survive
in the non-linear regime.  

\citet{Meheut2010,Meheut2012b} have also performed 3D simulations of the Rossby wave instability, but they have obtained a different evolution. 
They observed a strong axial recirculation inside the vortices. 
Although \citet{Meheut2010,Meheut2012b} considered a different framework (homentropic flow, whereas we consider adiabatic flow), 
we assume that the different evolution mainly comes from the different nature of the instability. In their case, the density bump is so large that it
induces a change of sign of $\kappa ^2$. Then, the disk also becomes unstable with respect to the centrifugal instability.
This instability induces strong axial recirculation. The formation of the Rossby vortices has probably been modified 
by this instability, and the axial flow observed in their simulation is the trace of the centrifugal
instability.  
\cite{MeheutYuLai2012} have also considered a case where $\kappa^2$ remains positive, so that the centrifugal instability
is absent. In that case,  they observed that the RWI remains mainly 2D and that the axial flow in the vortices
is low. However, they obtained a different  axial flow structure, which is probably related to their isothermal hypothesis.

\section{Stability and long-term evolution}
\label{sec:stability}

In the above section we have found that the RWI produces columnar vortices that can persist over many rotation periods. 
This implicitly means that the vortices are stable.  

This evolution is different from the evolution described by Barranco \& Marcus (2005). In the local shearing box approximation, 
they observed that columnar vortices are rapidly destroyed and   tend to reorganize in a layer of off-midplane vortices. 
It is not clear what causes this instability. We assume that it might be caused by elliptical instability that affects strained vortices
in the stratified and unstratified medium \citep{Lesur2009}. 
In the following section, we show that this instability can indeed destroy vortices within our framework.

\subsection{Elliptical instability}

The elliptical instability has been studied in the context of vortices in accretion disks by \cite{Lesur2009}.
They obtained with a local stability approach a condition for the instability of Kida-like vortices that depends on the horizontal aspect ratio of the vortices $\chi$ and on the stratification strength $N/S$ of the flow.

In this section, we analyze the conditions under which the instability  develops and discuss 
why the elliptical instability was not observed in the simulation presented in the previous section.

The high resolution required for the simulations was achieved by reducing the computational domain to a periodic box in the azimuthal direction; this is a simple way to avoid prohibitive increases of the CPU time. 
 The box size was ($6<r<9$ , $0<\theta<2\pi/10$ , $0<z<1$), and we used a resolution 
 of  $200\times100\times100$ in the $r$, $\theta$, and $z$ directions 
{(that is $33\times10\times50$ in number of cells per scale height).}

\begin{figure}
\begin{center}
\includegraphics[width=9cm]{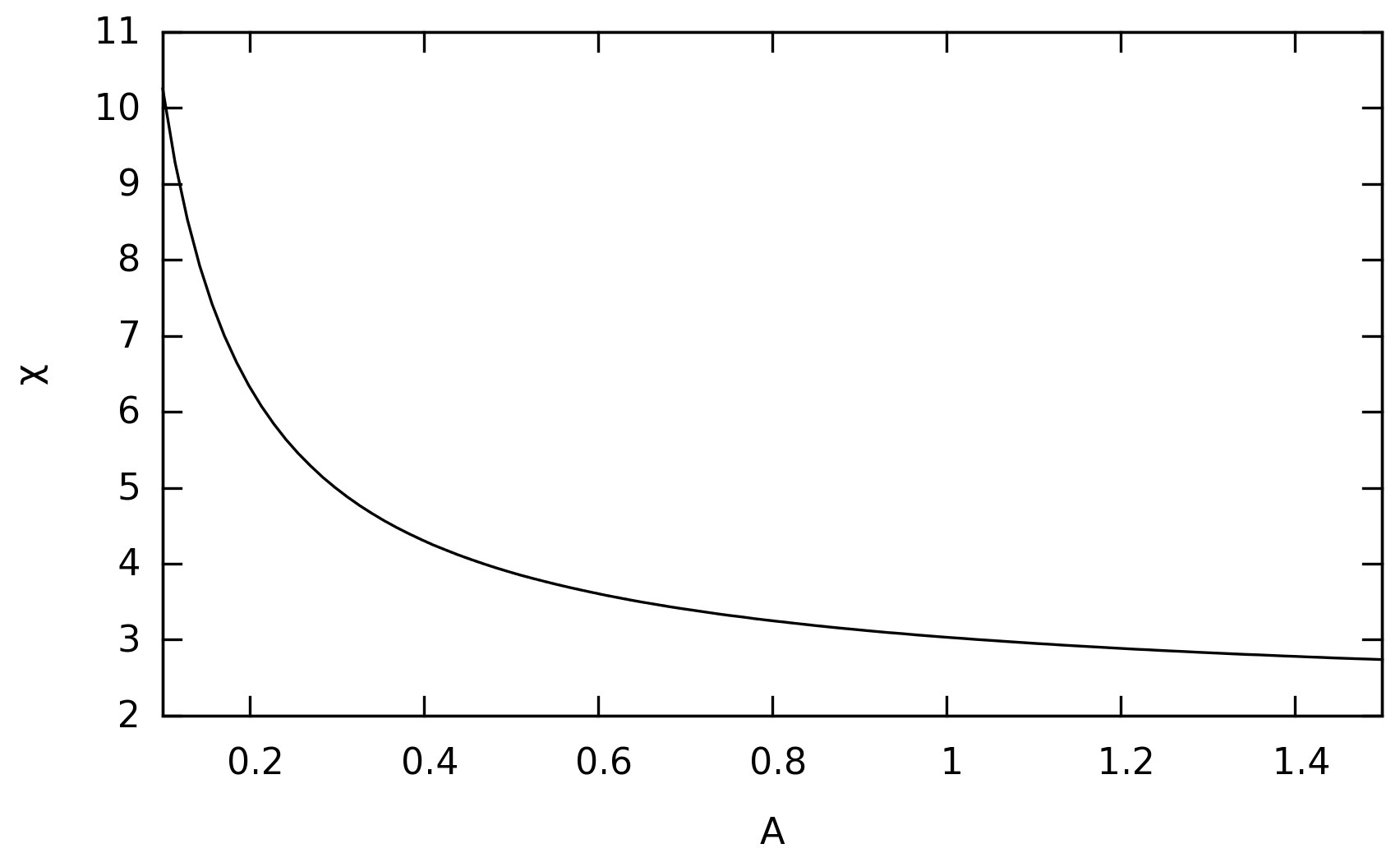}
 \caption{  { Theoretical prediction of the aspect ratio of the formed vortex as a function of the initial amplitude of the bump ($\sigma=0.3$, $r_b=7.5$, $p=1.5$, $q=0.5$, $M_{A_0}=25$). For these parameters, the basic flow is centrifugally unstable for $A>0.334$.}}
\label{chi-a}
\end{center}
\end{figure}

 We considered two different cases: (i) disks without vertical stratification, and (ii) disks with the same stratification as in Sect. \S \ref{sec:rossby}.
 For each case, different values of the aspect ratio were tested.  
 The aspect ratio was varied by changing the amplitude $A$ of the annular density bump used to trigger the RWI. 
 From the expression of the velocity field given in \S  \ref{sec:initial}, we can obtain an estimate of the maximum
 vorticity in the annular bump, and thus of the resulting vortex formed by the RWI,
 \begin{equation}
{\omega_{z_0}} \simeq -{r_b ~T_D(r_b) \over  \sigma^2  v_D(r_b,0)}{A \over 1+A} .
\end{equation}
The aspect ratio $\chi$ obtained from this formula { and equation (\ref{exp:Kida}) is plotted in Fig.~\ref{chi-a}. As observed in 
Fig.~\ref{fig:aspectratio}, this theoretical prediction provides a good approximation of the measured aspect ratio in the centrifugally stable regime.}
  
{Strong bump amplitudes $A$ are expected to produce vortices with a low aspect ratio, but they also imply negative values of $\kappa_0^2$. 
For this reason, low aspect ratio vortices are difficult to create. As soon as $A>A_c$, where $A_c \approx 0.334$ for the parameters of the simulation,
the centrifugal instability is active during the primary evolution of the bump. The Rossby wave instability is modified 
in such a way that  no columnar vortices are formed if the dynamics is not restricted to 2D.}
{ Three values of $A$ were considered in our study: $1.5$, $0.3$, and $0.2$, corresponding to the aspect ratios $2.5$, $5$, and $7$. 
For $A=1.5$ we obtained $\kappa_0^2 <0$ near the center of the bump.  The stratification effect of the low aspect ratio vortex 
that correspond to this value of $A$ is analyzed by a specific procedure.}

\subsubsection{Unstratified disks}

A disk without vertical stratification is obtained by replacing the gravitational potential $\phi=1/\sqrt{r^2+z^2}$ in equations (\ref{gov-eq1})-(\ref{gov-eq4}) by $\phi=1/r$. With this potential, the characteristics of the steady disk become  independent of $z$. 
In two dimensions, the Rossby wave instability is also active and gives rise to the formation of quasi-steady 2D Rossby vortices.

\begin{figure}
\begin{center}
\includegraphics[width=9cm]{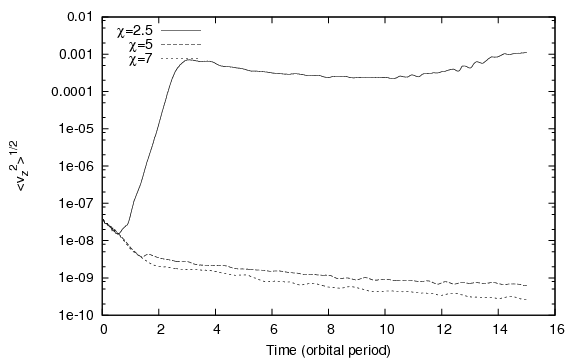}
 \caption{Time evolution of $\sqrt{<v_z^2>}$ for a vortex  in a unstratified disk as a function of its aspect ratio.} 
\label{vz}
\end{center}
\end{figure}

In the simulated azimuthal sector, a single 2D vortex is formed. Its 3D evolution is analyzed by considering this vortex in a 3D
environment, with periodic boundary conditions in the vertical direction. 
{Because the vortices are formed in 2D, the centrifugal instability is filtered out.   
{The centrifugal instability, which is characterized by the formation of axisymmetric rolls, is indeed inhibited when no axial displacement is allowed \cite[see for instance][]{Drazin_book}.}
Even for $A=1.5$, for which $\kappa_0^2<0$, a nice 2D vortex is 
obtained with an aspect ratio of 2.5, as predicted from the theory.}
As shown by LP09, the vortex aspect ratio is an important parameter for the stability. 
Figure \ref{vz} shows that different evolutions are observed for low and high aspect ratios. 
\begin{figure*}
\begin{center}
\begin{minipage}{.9\linewidth}
\includegraphics[width=4cm]{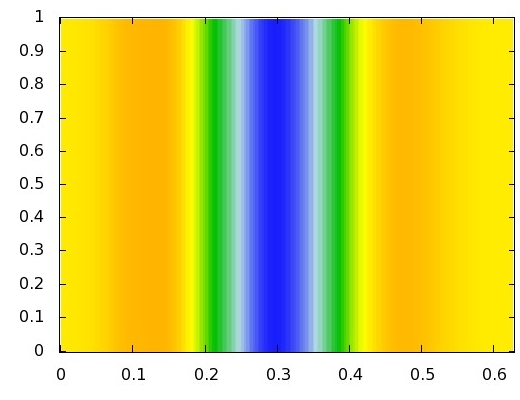}
\includegraphics[width=4cm]{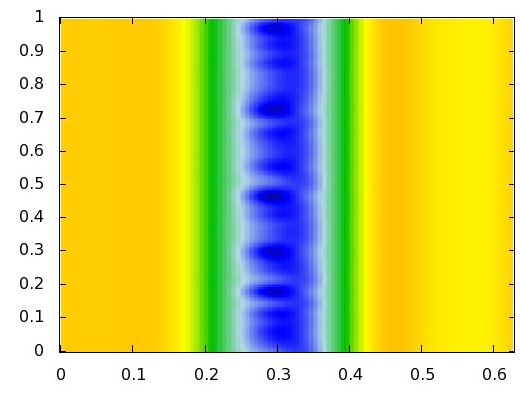}
\includegraphics[width=4cm]{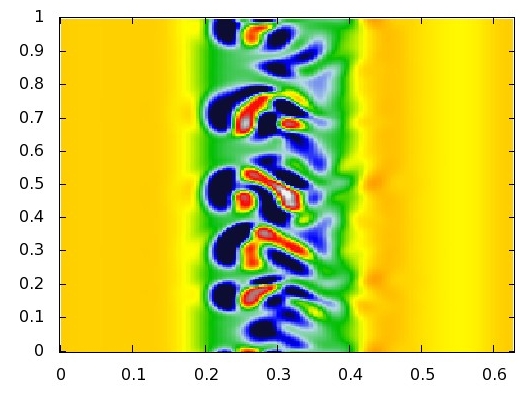}
\includegraphics[width=4cm]{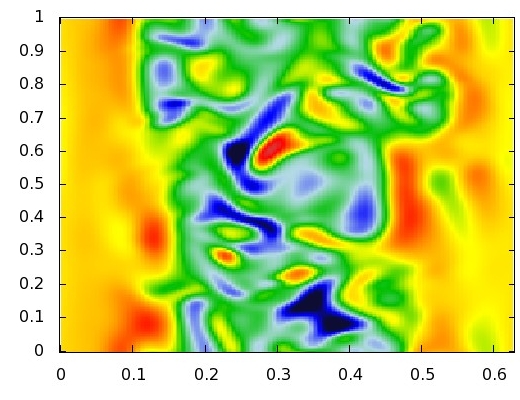}   
\includegraphics[width=4cm]{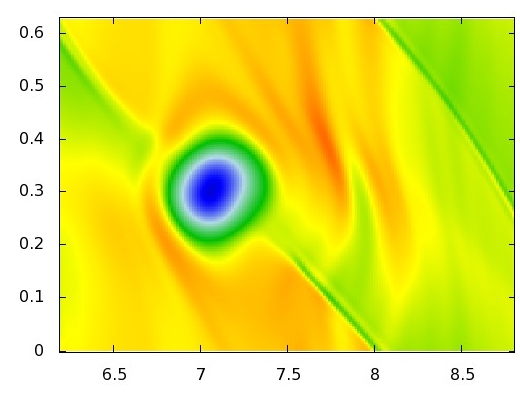}
\includegraphics[width=4cm]{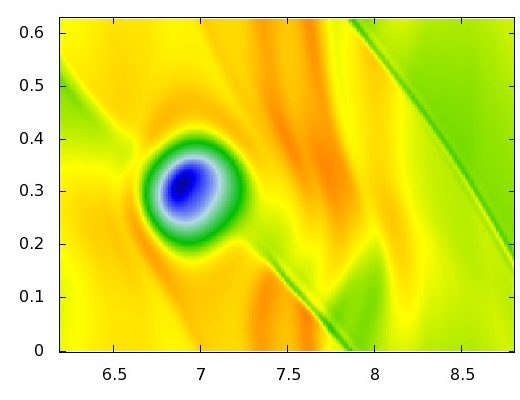}
\includegraphics[width=4cm]{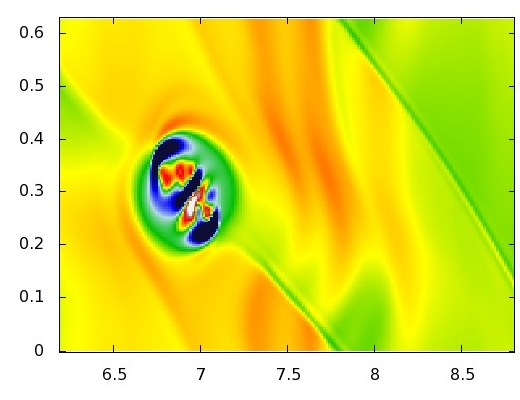}
\includegraphics[width=4cm]{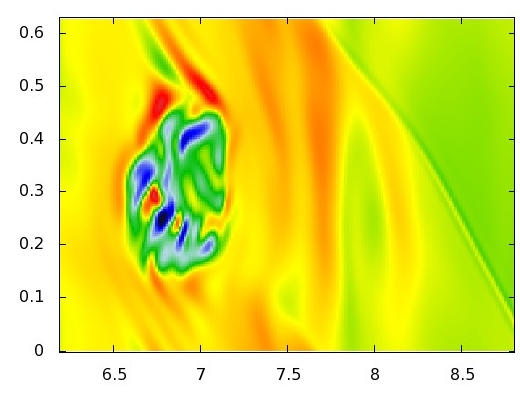}
\end{minipage}
\begin{minipage}{0.08\linewidth}
\includegraphics[height=6cm]{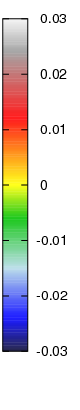}
\end{minipage}
\end{center}
 \caption{Evolution of a vortex of aspect ratio $\chi=2.5$ in an unstratified disk; from left to right: at initial time and after 2, 2.5, and 4 rotations. Top: vorticity in the $\theta$-z plane at the center of the vortex. Bottom: vorticity in the r-$\theta$ plane at level $z=0.5$.}
\label{ki=2.5}       
\end{figure*}
For low $\chi$, we observed the rapid growth of a 3D instability and the destruction  of the vortex in a few orbital periods.  
A typical evolution for $\chi = 2.5$ is illustrated in Fig.~\ref{ki=2.5}. This evolution is very similar to the 3D evolution shown in LP09. 
For  $\chi > 4$, no instability growth was observed: the vortex retains its columnar structure throughout the simulation. 
For example, for $\chi=7$, we found that the vortex persists for at least 200 rotations without being affected by any instability. 
These stable evolutions are  consistent with LP09 for  $4<\chi < 6$, but not for higher $\chi$.  LP09 predicted a destabilization of the vortex
for $\chi>6$ with a low growth rate around $10^{-2}$ the shear rate $s$. 
We have not seen such an instability. But one has to keep in mind that the theoretical predictions by LP09 are 
local and inviscid. Both the non-uniform character of the vortex and  the numerical  diffusion are expected to affect the 
instability growth rate.  
We verified that our results remained unchanged when the resolution was increased by a factor 2 in each direction.

\subsubsection{Stratified disks}

\cite{Lesur2009} have shown that vertical stratification affects the development of the elliptical instability.  
They have considered a uniformly stratified environment with a constant Brunt-V\"ais\"al\"a frequency $N$
for two values of the parameter $N/s$ ($N/s=0.1$ and $N/s=1$).
They have shown that in the regime of a low aspect ratio $\chi < 4$,  the local growth rate of the elliptical instability
is not modified by the stratification.  For $\chi >6$, the instability is still present, but with a smaller 
growth rate. For  intermediate aspect ratios $4<\chi <6$, they interestingly found a destabilization 
for  $N/s=1$. 

\begin{figure}
\begin{center}
\includegraphics[width=9cm]{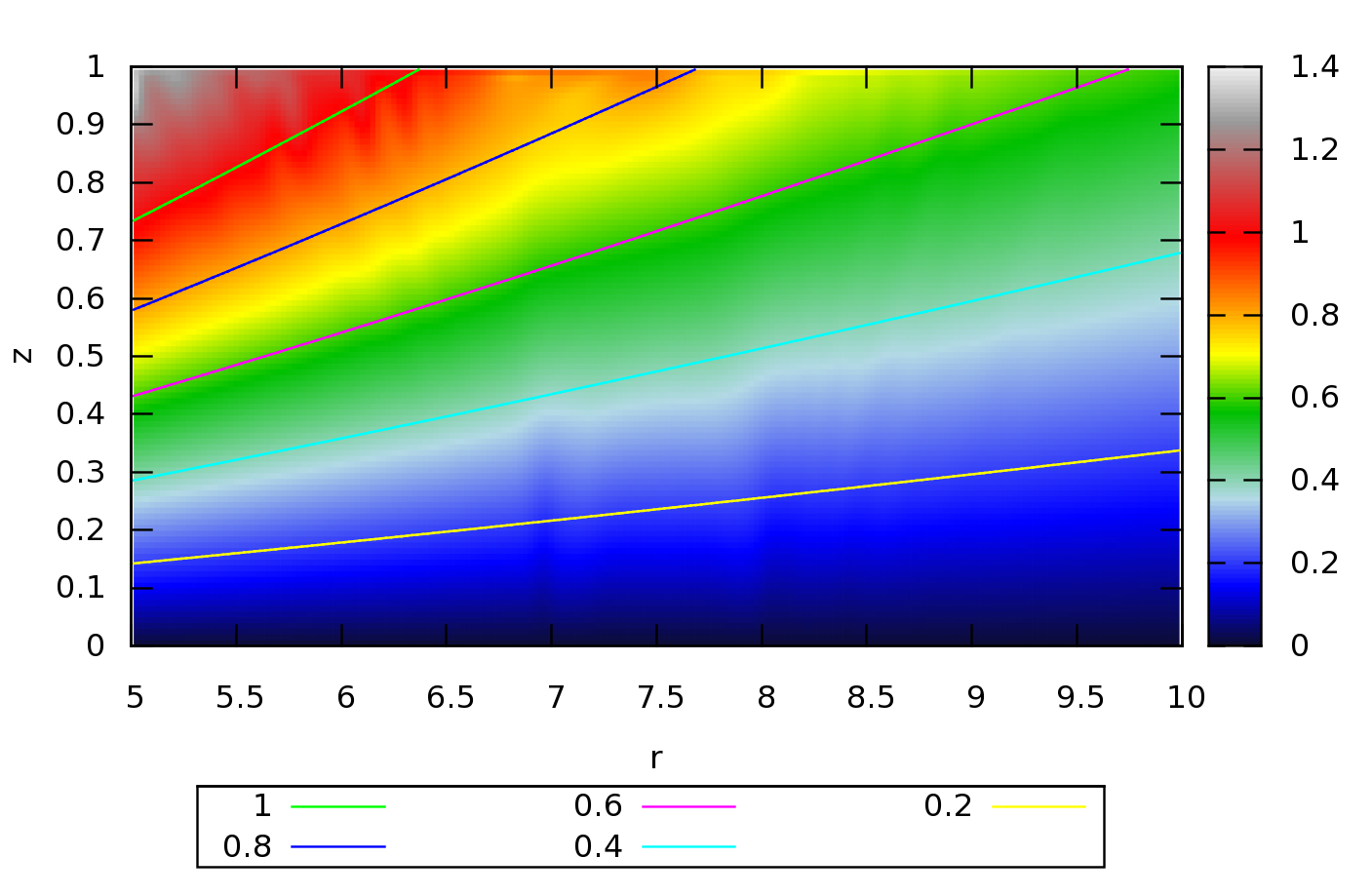}
 \caption{Stratification in the disk. The color map represents the iso-contours of $N/s$  in an angular sector going through the center of a vortex. The lines are the values obtained from Eq. (\ref{equ:N_s}) without vortices.}
\label{fig:N_S}
\end{center}
\end{figure}

 Here, the background disk is non-uniformly stratified with a parameter $N/s$ given by 
\begin{equation}
{N \over s} \approx {2 \over 3} \sqrt{{\gamma-1}\over \gamma} {z \over H_D(r)}~,
\label{equ:N_s}
\end{equation}
where the pressure scale height $H_D(r)$ has been defined in Eq. (\ref{exp:HD}).
The parameter $N/s$ therefore depends on the position.   
{Figure \ref{fig:N_S} shows the contours of $N/s$ in the presence of the vortices (along a cut crossing the vortex).  The lines are the contours obtained from Eq. (\ref{equ:N_s}).  This figure demonstrates that the vortices do not significantly modify the stratification and that 
$N/s$ is thus well predicted by Eq. (\ref{equ:N_s}).} 
This formula shows that $N/s$ vanishes in the midplane, but increases linearly with the vertical position in the disk. 
It is important to note that $N/s$ varies from 0 to 0.8 between the midplane and the disk edge 
at the location of the vortex.  We can therefore expect the top of the vortex to be affected by the stratification.

{ 
As above, we considered the evolution of bumps of  three different amplitudes $A=1.5$, $A=0.3$, and 
$A=0.2$.  
For the weak-amplitude cases ($A=0.3$ and $A=0.2$), we observed the rapid formation of a  columnar vortex 
 by the RWI  in a few rotations.  }
 
\begin{figure*}
\begin{center}
\includegraphics[width=4cm]{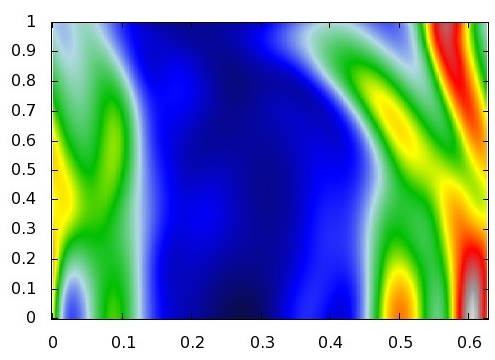}
\includegraphics[width=4cm]{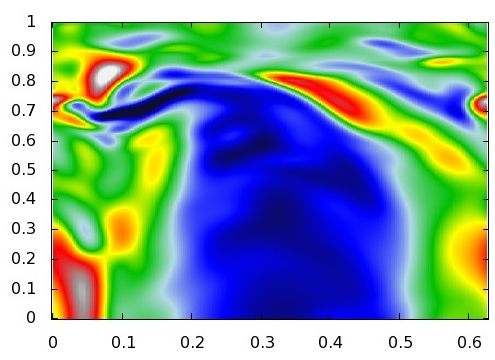}
\includegraphics[width=4cm]{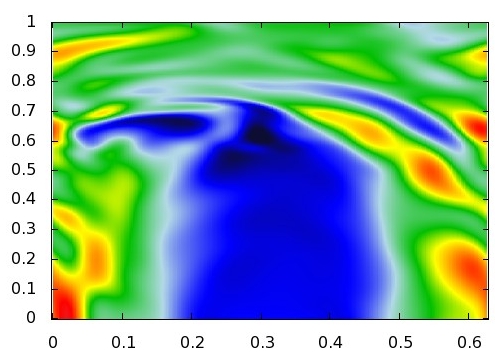}
\includegraphics[width=4cm]{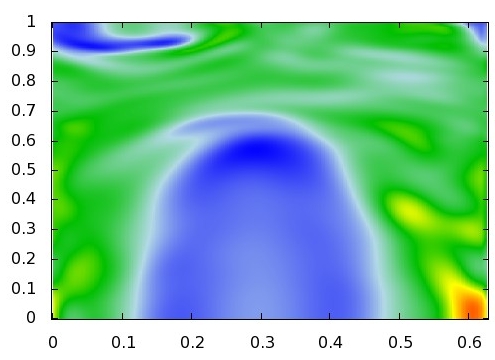}
\includegraphics[height=3cm]{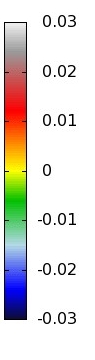}
\end{center}
 \caption{Vorticity in the $\theta$-z plane at the center of the vortex after 6 rotations, 15 rotations, 30 rotations, and 65 rotations.}
\label{stratif}       
\end{figure*}

 The amplitude $A=0.3$ gives rise to a vortex of aspect ratio $\chi=5$, as expected. 
 In this case, eventually, 
the higher part of the vortex ($z>0.7$) is destabilized and progressively disappears (see Fig.~\ref{stratif}). This occurs on a longer timescale than 
the rapid evolution observed in the unstratified case for  $\chi=2.5$.  After 60 rotations,  a stable but shorter vortex is obtained.   
This evolution is consistent with the theoretical prediction of LP09 for the elliptical instability.
In the weakly stratified region close to the midplane, the elliptical instability is not active for $\chi=5$. By contrast, 
at the top, where $N/s$ exceeds 0.5, the instability is probably present and responsible for the destruction of the
higher part of the vortex. 
Interestingly, we observe that the vortex apparently retains its shape in the region where  the elliptical instability is not active. 

For the vortex of aspect ratio $\chi=7$ obtained with $A=0.2$, we observed no destabilization.
This is consistent  with the results obtained in the unstratified case.
High aspect ratio vortices are indeed expected to be more stable in the presence of stratification. As they were found stable
in the unstratified case, the observed stable behavior in the stratified disk is not surprising. 
This stable behavior also explains why the elliptical instability was not observed in the 
simulation of the full disk. During the disk evolution, we notice indeed that the aspect ratio of the vortices remains higher than or equal to 7. 
The vortices are therefore always stable with respect to the elliptical instability  and retain their essentially 2D structure throughout the
simulation. 

{
For  $A=1.5$, the bump evolves in such a way that the vortices are destroyed during their formation.  The vortex of aspect ratio 2.5 that was obtained
in the unstratified 2D environment is not created in a stratified 3D disk. 
We assume that the reason is the presence of the centrifugal instability, which affects the vortex formation.}
{
Note also that the elliptic instability mechanism in low aspect ratio vortices is related to the centrifugal instability mechanism (LP09, Lyra 2013). Consequently, it is very difficult to determine whether the elliptic instability becomes active in this case.}

\begin{figure}
\begin{center}
\includegraphics[width=240pt]{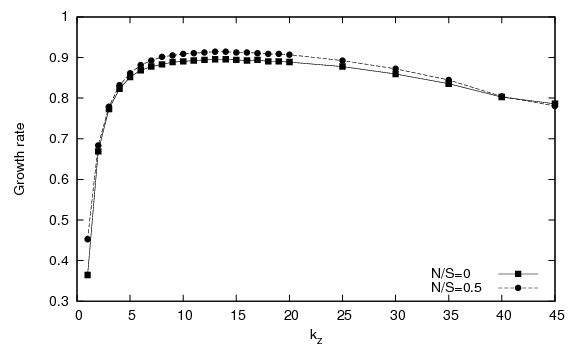}
\end{center}
 \caption{{Growth rate normalized by the local background angular velocity  of 3D perturbations in a vortex with an aspect ratio of 2.5 in an isothermal 
 stratified disk for  $N/s=0$ and $N/s=0.5$.}}
 \label{fig:ie_stratifie}
\end{figure}

{
To study the effect of the stratification on the dynamics of low aspect ratio vortices, we therefore used a different approach and considered isothermal disks with a uniform stratification along the height. 
For such a disk, a 2D simulation of a bump of amplitude $A=1.5$ 
is always possible and  leads to the formation of vortices of aspect ratio 2.5,  as before.  
These vortices can then be analyzed in a 3D environment with or without stratification. 
We considered two cases, $N/s=0$ and $N/s=0.5$.  Figure \ref{fig:ie_stratifie} shows the growth rate of 3D perturbations 
during the first instants as a function of the vertical wavelength $k_z$ in both cases. The growth rate here is normalized by  the background angular velocity $\Omega_D(r_b)$ at the position of the vortex. We see that the growth rate curves are almost the same in both cases, which demonstrates 
that the stratification effect on the stability characteristics of these vortices is very weak. 
This observation agrees with the results of LP09 who predicted no stratification effect when $\chi<4$.
The maximum growth rate also agrees with the value of 0.89 derived by LP09 for a Kida vortex of aspect ratio 2.5.
We also examined the non-linear evolution of the perturbation and observed that the vortex is rapidly destroyed in a way very similar to the case documented above in Fig.~\ref{ki=2.5}. }

We have mentioned the different evolution of the columnar vortices described in \citet{Barranco2005}.
In their simulation, the columnar vortex has a low aspect ratio $\chi=4$. In light of the above results, it is not 
surprising to observe the destabilization of the vortex because the elliptical instability is indeed active probably up to the midplane
for such a low aspect ratio.

\subsection{Migration of 3D vortices}

2D simulations show that vortices tend to migrate in Keplerian flows \citet{Li2000}. This migration process
 is due to the compression of the flow around the vortex and to the asymmetries in the position of the sonic lines and the associated density waves \citep{Paardekooper2010}. It also depends on the size of the vortex, on its aspect ratio and on the radial stratification of the disk \citep[e.g.,][]{Surville2012, Surville2013}.  Basically, it is observed that the higher the aspect ratio, the weaker the migration.  In this section, we present the first numerical study of the migration of 3D vortices. Our study focuses on vortices with high aspect ratios, first because we know from the previous section that vortices with $\chi >6$ are stable against the elliptical instability, but also because these vortices should migrate slowly due to their strong similarities with 2D vortices. 

Studying the migration of 3D vortices requires time-consuming long-term simulations. accordingly, as for the elliptical instability simulations, 
we simulated an azimuthal sector of the disk only.  A single vortex was formed from the RWI in a few rotations in the computational domain.  
 The dimension of the box was set up by the size and the aspect ratio of the vortex to be simulated. In practice we used a box size of ($5<r<10$, $0<\theta<\pi/2$, $0<z<1$) with a numerical resolution of $200 \times100\times100$  
{(or $33\times4\times50$ in number of cells per scale height)}.
 As for the elliptical instability, we changed the aspect ratio of the vortex 
 by changing the amplitude of the initial density bump. We fixed  $\sigma=0.3$ and took the values $A=0.2$, $A=0.15$, and $A=0.1$ to
  obtain vortices with an aspect ratio $\chi=7$, $\chi=8.5$, and $\chi=14$.  

The first steps of the evolutions are similar to what we described above. The RWI developed and after around 35 rotations, we obtained a single vortex 
in the computational domain that started to migrate toward the star. 
{\cite{Paardekooper2010} have noted that migration can stop close to a bump of the surface density, which has been interpreted by  \citet{Meheut2012b} as the reason why they did not observe migration. In our simulations, the initial density bump completely disappeared, when the vortices were formed. It therefore cannot disturb the vortex migration.}

\begin{figure}
\begin{center}
\includegraphics[width=240pt]{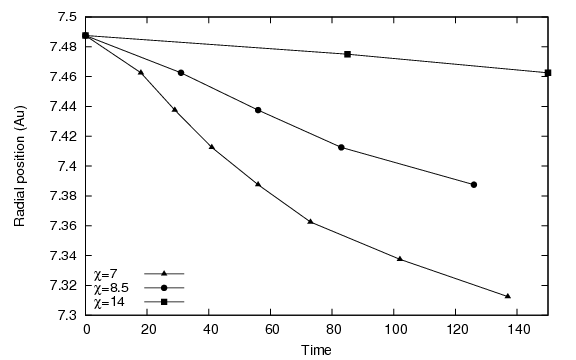}
\end{center}
 \caption{Radial position of the vortex as a function of time}
 \label{migration}
\end{figure}

The vortex survived more than 150 rotations and its structures were quasi-steady with only a weak time-evolution associated with the numerical diffusion. 
Figure \ref{migration} shows the radial position of each vortex as a function of time and for different aspect ratios. The migration rate over one rotation period was $1.5~10\textsuperscript{-3}$ AU for $\chi=7$,  $9.3~10\textsuperscript{-4}$ AU for $\chi=8.5$, and $1.3~10\textsuperscript{-4}$ AU for $\chi=14$. In non-dimensional values, the migration speed is $v_m$=3.1~10\textsuperscript{-5} $r_0\Omega_0$,  $v_m$=2~10\textsuperscript{-5} $r_0\Omega_0$, 
and $v_m$=3~10\textsuperscript{-6} $r_0\Omega_0$.

These values for the migration rate are of the same order of magnitude as in the 2D case \citep{Paardekooper2010, Surville2012, Surville2013}. This result is consistent with the structure of the vortices, which remained columnar and quasi-2D. In particular, migration seems to be independent of the disk thickness and no systematic deformation of the vertical vortex structure was observed in the radial direction. Of course we do not exclude that vortices with a different structure than ours may migrate in a different way.

\section{Conclusion}
\label{sec:conc}

Using fully compressible non-linear simulations in non-homentropic disks, we have found that the RWI can grow in 3D in nearly the same way as in 2D. Columnar vortices can be formed in a few  rotations and merge until a single vortex  remains in the disk. This vortex   
survives many rotation periods  and slightly drifts toward the star.  
Both the structure and the dynamics of the vortices were found to be mainly 2D.  
A small tilt of the vortex axis has nevertheless been observed. 

We showed that the linear phase of the RWI agrees with the results by \citet{Lin2013} both for the growth rate and 
the  structure of the unstable mode.  This result contrasts with the simulations by \citet{Meheut2010,Meheut2012b}, which showed the formation of vortices with a strongly z-dependent structure.  This difference is probably
mainly associated with the simultaneous presence of the centrifugal instability in their simulations. 
{In the centrifugally unstable case 
we simulated (for $A=1.5$),  the vortices were found to be destroyed during their formation.}  

The stability of the quasi-2D vortices with respect to the elliptical instability was explained by the high aspect ratio $\chi$ of the vortices. 
Using specific simulations, we showed that  vortices with $\chi >6$ are stable with respect to the elliptic instability in both a stratified disk and an unstratified disk. 
For intermediate values of $\chi$ ($4< \chi < 6$), we observed that the vortices survive close to the midplane, but are destroyed farther
away, at altitudes where the stratification is significant. We showed that this evolution is consistent with the theoretical results by \citet{Lesur2009} on the stratification effect on the elliptical instability. For a low aspect ratio $\chi < 4$, the vortices were shown to be strongly unstable. They  are rapidly destroyed in unstratified disks and isothermal stratified disks.   
These results can perhaps explain the rapid destruction of the columnar vortex of aspect ratio 4 
simulated by \citet{Barranco2005}. Here, we argued that the destruction of  2D vortices cannot be systematic 
and that the vortices probably remain columnar for a long time when their aspect ratio is sufficiently high, that is, when they are sufficiently weak. 

The migration process of these quasi-2D vortices was studied as well. 
We showed that the drift velocity decreases with the aspect ratio, as previously observed in 2D.

In actual protoplanetary disks, the initial overdensities necessary to trigger the RWI may be caused by an axisymmetrical accumulation of gas at the edge of the dead zone. As the density would increase gradually under the turbulent transport outside of the dead zone, we would expect the RWI to occur close to its critical stability condition, that is, for a low density bump amplitude. The vortices formed by the instability would therefore be weak and have a high aspect ratio. 
They would then be stable with respect to the elliptic instability and have a low migration speed. This means that these vortices
would survive a sufficiently long time far away from the star to be able to capture and confine large amounts of solid particles.
These weak vortices would therefore be good sites for the formation of planetesimals or planetary cores. 

The recent discovery of an asymmetry in the dust distribution in the disk of Oph IRS 48 by ALMA reported by \citet{vdMarel2013} and \citet{Armitage2013} has been interpretred as an assembly of particles captured by a large-scale Rossby vortex. According to the authors, this vortex has formed by an RWI occurring at the edge of the gap opened by a massive planet orbiting in the inner part of the disk. These observations indeed seem to confirm the efficiency of the capture in vortex mechanism pointed out by \citet{Barge1995} and open up the possibility to derive other observational constraints.

\begin{acknowledgements}
We thank the anonymous referee for his useful comments.
Computations were performed on a Bull multicore machine funded by the Aix-Marseille University and the Laboratoire d'Astrophysique de Marseille; the most demanding computations were performed on CURIE at TGCC (contract gen6832 2012/2013). This project was also supported by the French ``Programme National de Plan\'etologie" (PNP2012).  
This work has received support from the French National Research Agency under the A*MIDEX grant ANR-11-IDEX-0001-02, and is part of the LABEX MEC project ANR-11-LABX-0092.
3D visualisations were performed with \textit{glnemo2} developped at CeSAM - http://projets.lam.fr/projects/glnemo2.
\end{acknowledgements}

\nocite{*}
\bibliographystyle{aa}
\bibliography{3D-Rossby-8}

\end{document}